# Direct Observations of Formation and Propagation of Subpolar Eddies into the Subtropical North Atlantic

Amy S. Bower[a*], Ross M. Hendry[b], Daniel E. Amrhein[c], Jonathan M. Lilly[d]

[a] *Department of Physical Oceanography, Woods Hole Oceanographic Institution, Woods Hole, MA 02543, United States*
[b] *Ocean Sciences Division, Fisheries and Oceans Canada, Bedford Institute of Oceanography, Dartmouth, Nova Scotia, Canada B2Y 4A2*
[c] *WHOI/MIT Joint Program, Woods Hole Oceanographic Institution, Woods Hole, MA 02543, United States/Massachusetts Institute of Technology, 77 Massachusetts Avenue, Cambridge, MA 02139, United States*
[d] *NorthWest Research Associates, P.O. Box 3027, Bellevue, WA 98009, United States*

________________

[*] *Corresponding Author at: Department of Physical Oceanography, Woods Hole Oceanographic Institution, Woods Hole, MA 02543, United States*
*E-mail address: abower@whoi.edu*

ABSTRACT

Subsurface float and moored observations are presented to show for the first time the formation and propagation of anticyclonic submesoscale coherent vortices that transport relatively cold, fresh subpolar water to the interior subtropical North Atlantic. Acoustically tracked RAFOS floats released in the southward-flowing Western Boundary Current at the exit of the Labrador Sea reveal the formation of three of these eddies at the southern tip of the Grand Banks (42°N, 50°W). Using a recently developed method to detect eddies in float trajectories and estimate their kinematic properties, it was found that the eddies had average rotation periods of 5–7 days at radii of 10–25 km, with mean rotation speeds of up to 0.3 m s$^{-1}$. One especially long-lived (5.1 months) eddy crossed under the Gulf Stream path and translated southwestward in the subtropical recirculation to at least 35°N, where it hit one of the Corner Rise Seamounts. Velocity, temperature and salinity measurements from a nine-month deployment of two moorings south of the Gulf Stream at 38°N, 50°W reveal the passage of at least two eddies with similar hydrographic and kinematic properties. The core temperature and salinity properties of the eddies imply their formation at intermediate levels of the Labrador Current south of the Tail of the Grand Banks. These observations confirm earlier speculation that eddies form in this region and transport anomalously cold, low-salinity water directly into the subtropical interior. Possible formation mechanisms and potential importance of these eddies to interior ventilation and the equatorward spreading of Labrador Sea Water are discussed.

# 1. Introduction

Submesoscale coherent vortices (SCVs) are long-lived, small (compared to the mesoscale), energetic, anticyclonic eddies that have been observed to transport volumes of anomalous water a thousand kilometers or more from their formation sites (see McWilliams, 1985 for a review). The first observation of an SCV in the ocean was described by McDowell and Rossby (1978) in the western North Atlantic near the Bahamas. They discovered an anticyclonic, intra-thermocline lens with a core of high-salinity water characteristic of Mediterranean Outflow Water (MOW) from the far eastern North Atlantic. This discovery of a new mechanism for large-scale transport and mixing of water properties prompted a search for more Mediterranean eddies, or Meddies, in the eastern North Atlantic. Many meddies have subsequently been found and studied in detail (see e.g., Armi et al., 1989; Richardson et al., 1991; Richardson et al., 2000), and their formation from the Mediterranean Undercurrent along the continental slope of the Iberian Peninsula has been observed and modeled (Pichevin and Nof, 1996; Bower et al., 1997; Jungclaus, 1999; Cherubin et al., 2000; Serra et al., 2005; Serra and Ambar, 2002). As predicted by McDowell and Rossby (1978), it has been estimated that Meddies are responsible for a significant fraction of the salt flux away from the Mediterranean (Richardson et al., 1989; Arhan et al., 1994; Maze et al., 1997), evidence that SCVs can have an important role in re-distributing water properties over large spatial scales. Ironically, Prater and Rossby (1999) have updated the proposed origin of the first meddy, arguing that it formed in the Northwest Corner region of the North Atlantic Current.

McWilliams (1985) proposed that SCVs mainly form when diapycnal mixing creates a volume of water with weak stratification that subsequently spins up into an anticyclonic eddy by geostrophic adjustment when the volume is injected into a more stratified fluid. This mechanism has been demonstrated in laboratory models (see e.g., Hedstrom and Armi, 1988). Alternatively, D'Asaro (1988) proposed that anticyclonic vorticity is generated by frictional torque in a wall-bounded jet which then separates from the coast at a sharp corner and forms an eddy. He showed with observations that Beaufort Sea SCVs could form in such a manner. Similarly, Prater (1992) and Bower et al. (1997) argued that there is sufficient horizontal shear in the Mediterranean Undercurrent to form a meddy with the observed relative vorticity without the need to collapse a weakly stratified volume of MOW. Bower et al. (1997) further showed with subsurface floats that some meddies do form where the boundary current tries to negotiate a sharp corner in the continental slope of the Iberian Peninsula, where the radius of curvature is less than the local radius of deformation.

Most SCVs in the ocean are difficult to detect and track because of their relatively small horizontal scale (<50 km radius) and general lack of a measurable surface expression. Most have been discovered by accident. A subthermocline SCV with a cold, low-salinity, high-dissolved-oxygen core, called D1, was found near 31°N, 70°W during the mesoscale-focused Local Dynamics Experiment (LDE) and surveyed extensively with SOFAR floats and with expendable hydrographic and current profilers (Lindstrom and Taft, 1986; Elliott and Sanford, 1986a, b). D1's lens-like structure was centered at 1500 m, and density perturbations extended from 1000 m to below 3000 m. Direct velocity observations revealed an eddy core in solid-body rotation, with a maximum anticyclonic azimuthal speed of 0.29 m s[-1] at a radius of 15 km from the eddy center, decaying exponentially to background levels beyond a radius of 25 km. Based on the large-scale salinity distribution and the basic pattern of circulation in the western North Atlantic, they speculated that D1 formed in the vicinity of the Grand Banks of Newfoundland, was first carried eastward by the Gulf Stream, and then southwestward in the subtropical recirculation (see Figure 1 for locations and major circulation features). D'Asaro (1988) also speculated that SCVs with a core of cold, low-salinity subpolar water could form at the southern tip of the Grand Banks (the Tail of the Grand Banks, or TGB) via the mechanism he proposed for the formation of Beaufort Sea eddies.

Here we present results from eddy-resolving subsurface float trajectories and time series of currents and temperature/salinity from a moored array that show directly for the first time the formation and propagation of cold, low-salinity SCVs from the TGB to the interior subtropical North Atlantic, confirming the earlier speculation about the TGB as an eddy formation site. Here the Western Boundary Currents (WBC[1]) of the subpolar gyre, transporting the recently ventilated components of North Atlantic Deep Water at depth and waters of polar origin in the upper layer, encounter a sharp clockwise bend in the bathymetry of the middle and upper continental slope and a major extension of the lower continental slope called the Southeast Newfoundland Ridge (SENR; Figure 1). The eastward-flowing Gulf Stream is negotiating the same bathymetry as some fraction of its transport turns sharply northward as the North Atlantic Current. Traditionally, it had been thought that NADW makes the turn around the TGB as a continuous current to follow the continental slope equatorward in the DWBC. Recent Lagrangian observations revealed however that there are also interior southward pathways through the subtropics, at least for Labrador Sea Water (LSW) in the depth range 700–1500 m (Bower et al., 2009; Bower et al., 2011). These pathways are consistent with eddy-driven recirculation gyres adjacent to the Gulf Stream and North Atlantic Current (NAC) (Lozier, 1997; 1999; Gary et al., 2011). As will be shown in the following sections, at least some of this interior southward transport of subpolar waters into the subtropical region is accomplished by SCVs.

---

[1] We use the name Western Boundary Current here to refer to the collection of currents transporting subpolar and Nordic Seas waters southward along the western boundary, including the Deep Western Boundary Current and the baroclinic and barotropic branches of the Labrador Current (Lazier and Wright, 1993).

In section 2 below, the data sets and methods for identifying eddies in the float trajectories are described. In section 3.1, three float-detected eddy formation events are documented and the kinematic properties of the eddies are estimated. Two eddies observed to pass a moored array south of the TGB and Gulf Stream are described in section 3.2. Some clues regarding the formation process of these eddies are discussed in section 4 and the results are summarized in section 5.

# 2. Data

## 2.1 RAFOS Float Data

The float trajectories analyzed here are a subset of 59 isobaric RAFOS floats (Rossby et al., 1986) that were released in the western boundary current near 50°N (see Figure 1 for release location) as part of a larger study of LSW spreading pathways called *Export Pathways from the Subpolar North Atlantic*. The *ExPath* floats were ballasted for either 700 or 1500 dbar, the nominal core depths of Upper LSW (ULSW) and Classical LSW (CLSW). Floats were released nominally in groups of six every three months for three years starting in July 2003. They were tracked using an international array of sound sources. Position fixes and temperature and pressure were measured once daily by the floats and stored internally for the two-year drifting mission, after which the floats surfaced and returned all their data via Service ARGOS. Pressure and temperature measured by the floats are accurate to within ±5 dbar and ±0.005°C, respectively. More details on the full RAFOS float data set can be found in Furey and Bower (2009) and Bower et al. (2011).

## 2.2. Moored Measurements: BIO-50W Moored Array

The mooring data discussed here came from the East and South sites of a three-mooring triangular array with a common 50 km separation centered at 38°N and 50°W in water depths just over 5200 m (Figure 1). This will be referred to as the BIO-50W moored array in the discussion that follows. All moorings were equipped with Aanderaa current meters measuring pressure, temperature, conductivity, rate, and direction at nominal 1000, 1200, and 1500 m depths from 04 December 1978 to 14 September 1979 for a maximum of 287 days. These moorings also had measurements at 4000 m and deeper that are not discussed here. All upper level North site instruments were lost when the mooring parted sometime during the deployment. Details can be found in Hendry (1985). The results described in this paper are based on low-pass filtered data using two applications of a second-order Butterworth filter (forward and backward to preserve phase) with a cutoff period of 18 hours. The results were subsampled at common 6-hour intervals. The data set is less than ideal: there are periods of missing rate data at 1200 m at the East site and at 1200 m and 1500 m at the South site, and the 1200 m conductivity data at both East and South sites are contaminated by extreme noise.

Neither the float nor the mooring observations described here were made with the idea of observing submesoscale eddies forming at the TGB and propagating into the subtropical interior.

As a result neither data set is ideal for studying these SCVs. However, the observations reveal without a doubt that SCVs with cores of relatively cold, fresh subpolar water form at the TGB and enter the subtropical circulation, exposing a new mechanism for ventilation of the subtropical interior and motivating a future dedicated study of these eddies.

## 2.3. Extracting Eddy Signals from RAFOS Float Trajectories

We use a relatively new method, built on the continuous wavelet transform, for estimating the properties of the eddies observed with the RAFOS floats. The basic idea of the analysis method is to begin with a mathematical model for the displacement signal of a Lagrangian float orbiting the center of an eddy, and then to estimate this signal directly by identifying the "best fit" parameters using a procedure known as wavelet ridge analysis. This leads to the estimation of four important time-varying quantities that describe the oscillatory motion of an eddy-trapped float: a radius $R(t)$, a velocity $V(t)$, a frequency $\omega(t)$, and a measure of shape $\xi(t)$ related to the eccentricity. The estimation of an oscillatory signal in this manner is the opposite of the common approach of defining the eddy currents to be the residual after subtracting an estimated background flow. The method details have been examined in a series of papers (Lilly and Olhede, 2009a, b, 2010, 2011) based on an earlier prototype study by Lilly and Gascard (2006). The essential aspects of the method are discussed in the Appendix, with a focus on issues of interpretation and practical implementation. The wavelet ridge analysis is carried out using routines from a freely distributed software toolbox (Lilly, 2011).

# 3. Results

## 3.1 Eddy Formation at the Tail of the Grand Banks Observed with RAFOS Floats

Three eddy formation events were observed with RAFOS floats at the TGB between October 2005 and August 2006, referred to here as eddies A, B and C. Eddy formation is evident when a float trajectory changes from a smooth curve to a cusped or epicyclical motion indicative of the float looping in a translating eddy. Eddy A was observed with a 1500-dbar RAFOS float, and eddies B and C were revealed by 700-dbar floats. Each event is described in detail below. The dates of eddy observations are given in Table 1a and the kinematic properties of each eddy are summarized in Table 1b.

### 3.1.1. Eddy A

The first eddy to form at the TGB in the *ExPath* RAFOS float trajectories was captured by the 1500-dbar float 664. Its trajectory, color-coded by float temperature, is plotted in Figure 2a superimposed on the climatological mean temperature at 1500 meters from HydroBase (Curry, 2002; http://www.whoi.edu/hydrobase). This float was released in May 2005 over the 1800-m isobath near 50°N. It drifted equatorward over the continental slope, passing around Flemish Cap on its way to the TGB, with an average speed of 0.12 m s$^{-1}$. In October 2005, the float track showed several shallow cusps as it diverged from the 2500-m isobath at the TGB and drifted into deep water. The track became irregular during November due to poor sound source reception, then clear cusped motion appeared again around 10 December 2005, indicative of a rapidly translating anticyclonic eddy. Consistently closed loops developed around 1 February 2006, and continued until the eddy drifted very close to one of the Corner Rise Seamounts near 35°N, at the end of March 2005, at which point float looping stopped abruptly. After looping stopped, float 664 continued drifting generally southwestward.

Figure 2a and the time series of float temperature and mean climatological temperature along the float path, Figure 2b, show that Eddy A transported a core of relatively cold water (as much as 0.7°C colder than climatology) into the warmer subtropical region over its five-month lifetime. Float temperature was low and relatively constant while the float was drifting southward in the boundary current. It fluctuated between 3.4 and 3.6°C as the float left the slope, then was relatively constant (low temperature variance) at 3.65°C through early December, when the track was not well-resolved. Float temperature increased to 3.8°C over seven days during the first half of December. Below we will show that this coincides with when Eddy A hit the Gulf Stream and is the signature of the eddy being subducted under the Gulf Stream along isopycnals. Float temperature remained nearly constant at 3.8°C until the end of March 2006, when temperature again increased abruptly by 0.5°C and float looping stopped.

The extended periods of low float temperature variance suggest that the float was trapped in the relatively homogeneous eddy core, a common feature of SCVs due to homogenization of properties within the radius of maximum azimuthal velocity (McWilliams, 1985). The abrupt increases in temperature and temperature variance when the float looping stopped suggest that the float had left the eddy core. Float temperature remained about 0.2°C colder than the mean background temperature for a few months after looping stopped, possibly indicating that the float was embedded in some remnants of the eddy core water.

Application of the wavelet ridge analysis (see Appendix) to the 664 float trajectory first identifies an eddy, defined as when the geometric mean radius R first becomes > 0, on 22 October 2005, just as the float is diverging from the continental slope at the TGB, Figure 3. After 14 days, the wavelet analysis temporarily "loses" the eddy signal when the track quality is marginal (see below). On 22 November 2005, an eddy is again detected in the float trajectory, and is continuously tracked by the wavelet analysis until the float stopped looping at the seamounts at the end of March 2006.

The eddy properties determined by the wavelet analysis are plotted in Figure 4 as a function of time. As demonstrated with an idealized eddy example in the Appendix, one or two rotation periods are required before the method can be used to accurately estimate the eddy properties. Therefore, in this and subsequent plots of eddy properties, dotted curves indicate two rotation periods at the ends of each eddy segment. These points will not be used to estimate average eddy properties. Note however that an eddy can be detected by the analysis technique *before* reliable estimates of the eddy's properties can be obtained (see Appendix). Start and end dates for eddy detection and for estimating kinematic properties are given in Table 1a.

Due to the short length of the first eddy segment in late October (blue lines) and the poor quality of the track during November, the analysis method does not give reliable estimates of eddy properties during this time. However, the time-of-arrival (TOA) record from one continuous sound source signal, Figure 5, provides strong evidence of the onset of oscillatory motion (in the form of cusps in the TOA) at the TGB in October and its persistence until the track quality improves in December. The low temperature variance through November, Figure 4d, also indicates the float was in an eddy core throughout this time period. The average period of the oscillations in travel time is about five days.

During the period of reliable estimation of eddy properties (solid green curves in Figure 4), the geometric mean radius R gradually increased from about 10 to 22 km. The geometric mean looping speed V increased (in the anticyclonic, or negative, direction) from about $-0.18$ m s$^{-1}$ to a maximum of about $-0.37$ m s$^{-1}$. These changes are linearly correlated and represent a nearly constant looping period of $5.3\pm0.3$ (standard deviation) days. Average eddy properties from this ~4-month period are listed in Table 1b.

A plot of V as a function of R, Figure 6, confirms the linear relationship between V and R and that the float was embedded in an eddy core in near solid-body rotation during the stable period (solid green symbols). The relative vorticity of the solid body core is estimated for the stable period by a linear fit is $\zeta=-2.8 \times 10^{-5}$ s$^{-1}$, giving a vorticity Rossby number of 0.3. Such a large Rossby number implies that the eddy is not in geostrophic balance and that centripetal acceleration is important. The points from the first eddy segment (open blue circles, which correspond to dashed lines in Figure 4) fall close to the same line, but as shown in the Appendix, Figure A2, this is an artifact of the analysis.

Superposition of the float track on maps of absolutedynamic topography (ADT) from Aviso (http://www.aviso.oceanobs.com), where the merged, 1/3°x 1/3° gridded sea level anomaly fields have been combined with a mean  dynamic topography, illustrate how the eddy and float 664 crossed into the subtropical recirculation, Figure 7. The paths of the meandering Gulf Stream and North Atlantic Current are evident from the strongest gradients in the dynamic topography. The lowest sea surface height (SSH) in the vicinity of the TGB is associated with a trough between the eastern slope of the Grand Banks and the NAC. This minimum in SSH divides the southward surface flow associated with the Labrador Current over the middle and upper continental slope from the northeastward flow of the NAC farther offshore (Rossby, 1996; Fratantoni and McCartney, 2010).

The sequence of ADT images shows that as float 664 was drifting southward over the eastern slope of the Grand Banks, the trough of low SSH was expanding southward also and wrapping around the TGB. Float speed was increasing from near zero to about 0.15 m s$^{-1}$ just before the eddy formed (Figure 4e). The extension of the low SSH trough continued until 23 November 2005, when an isolated area of low SSH appeared to break off from the 'nose' of the trough while the main body of the trough retreated back to the north. Float 664 and Eddy A drifted southward along the western edge of the cut-off low, and from that time until Eddy A collided with the Corner Rise Seamounts, the direction of the eddy translation was consistent with the direction of the surface geostrophic flow indicated by the ADT gradient. The translation of Eddy A increased dramatically in the first half of December to 0.30 m s$^{-1}$ (Figure 4e), apparently when it was entrained into the Gulf Stream. Around 1 February 2006, the translation rate of the eddy decreased and it turned more southward, then southwestward, apparently around the eastern terminus of the subtropical recirculation. The relatively slow southwestward drift of the eddy at 0.05–0.10 m s$^{-1}$ continued until the eddy reached the seamounts at the end of March. Eddy A's southwestward drift was consistent with the direction of the surface geostrophic flow between an area of high ADT to the northwest of Eddy A and low ADT to the southeast. Eddy rotation period remained remarkably constant throughout the periods of both fast and slow eddy translation, Figure 4c, e.

Figure 8 shows an expanded view of the interaction between Eddy A and the Corner Rise Seamounts, and Figure 9 shows the silhouette of the seamount chain as seen from the north and the position of the eddy when float 664 stopped looping. The seamounts obviously represent a major obstacle course for SCVs approaching from their formation site at the TGB. The abrupt cessation of looping, and accompanying increase in float temperature when the eddy passed close to the seamount are compelling indicators that the eddy collided with the topography. With this single trajectory, it cannot be determined if Eddy A was completely destroyed by the collision, but a major disruption of the eddy circulation at the radius of float looping clearly occurred. Similar behavior was observed when meddies being tracked by RAFOS floats collided with seamounts in the eastern North Atlantic (Richardson et al., 2000), a process that has been studied with numerical and laboratory models (Cenedese, 2002; Wang and Dewar, 2003; Adduce and Cenedese, 2004).

### 3.1.2 Eddy B

Eddy B was observed with *ExPath* RAFOS float 581, which was ballasted for the level of ULSW in the WBC at 700 dbar. Figure 10a shows the first 200 days of its trajectory with the climatological mean temperature at 700 dbar from HydroBase. Float 581 was deployed during November 2005 and temporarily anchored to the seafloor until 15 February 2006, when it was programmed to release its anchor and begin its drifting mission. It was released over the 1400-m isobath, the most inshore launch site for the 700-dbar floats in *ExPath*. Like most of the floats released at this site, float 581 drifted southward through Flemish Pass, the 1100-m deep channel between the Grand Banks and Flemish Cap (Figure 1; see also Bower et al., 2011). It continued drifting southward over the upper continental slope east of the Grand Banks, with a mean speed of 0.17 m s$^{-1}$.

Float temperature, Figure 10a, b, was low and constant within 0.3°C while the float was drifting southward in the boundary current. Near the end of June 2006, float 581 diverged from the 1000-m isobath and crossed the slope without rounding the TGB. Five anticyclonic cusps in the trajectory followed over the next 30 days, indicating that the float was trapped in a translating anticyclonic eddy. For the first 18 days of looping, the eddy was advected southeastward along the flank of the SENR, and float temperature remained low and relatively constant at about 3.8°C. Then float temperature increased to about 4.5°C in several irregular steps during the rest of the float looping. Near the end of July, looping ceased and the float drifted slowly southward. In early August, float temperature shot up by about 3°C within two days, consistent with a crossing of the Gulf Stream.

Even with only five apparent rotations of Eddy B, the wavelet ridge analysis successfully detects it on 22 June 2006, Figure 11. However, with only five apparent rotation periods, estimation of the eddy properties with the same analysis method as used for float 664 is limited to a short 10-day time period in the middle (not shown). Average eddy properties during that time are listed in Table 1b. This example is shown here mainly to illustrate another eddy formation event, even though its short life (32 days) prevents detailed analysis of its properties.

The large-scale context provided by the maps of ADT, Figure 12, shows that float 581 was drifting rapidly along the upper continental slope inshore of the sea surface trough as it approached the TGB, 14–28 June 2006. Where the float and eddy diverged from the slope, the ADT contours are closely spaced and perpendicular to the bathymetry, indicating a strong offshore surface current between the end of the cold trough and a sea surface high (Gulf Stream meander or warm core ring) to the west. Over the next month, the float and eddy translated parallel to the SSH contours as they more or less follow a weaker extension of the ADT trough that was directed generally southeastward from the TGB. Up until 12 July, the eddy trajectory and the surface current are both directed along the flank of the SENR. Subsequently the eddy trajectory turned abruptly eastward, still following the orientation of the surface flow, and float temperature increased, Figure 10b. About 12 days later, looping stopped, but temperature remained around 4.5°C until the float intersected the Gulf Stream on 14 August.

### 3.1.3 Eddy C

The last eddy observed with the ExPath RAFOS float data set was different from the previous two eddies in that it translated westward rather than southward after separation at the TGB. It is also unique in that two 700-dbar floats (680, 582) were trapped in its core, while a third (586) drifted westward with Eddy C at its periphery. Eddy C's life history is illustrated in Figure 13a using float 680. This float was released in April 2006 over the 1800-m isobath near 50°N and drifted through Flemish Pass on its southward path toward the TGB. Float temperature, Figure 13a, b, matched the mean background until mid-August 2006, when the float began to cross the continental slope and the climatological temperature gradient at the TGB. Float speeds reached nearly 0.30 m s$^{-1}$ just as the first eddy motion was detected. After a few anticyclonic cusps, the trajectory formed closed loops and the new eddy translated westward for 4.1 months, measuring temperatures consistently about 1.5°C lower than the mean background temperature. When float looping stopped in late December 2006 and the float accelerated toward the south, float temperature increased abruptly by 4°C, indicating that the float had left the eddy.

Figure 14 shows that the wavelet ridge analysis detects two eddies (blue and green curves) in this one trajectory because there is an apparently rapid change in looping period of the float in September. The time series of TOA for sound signals from source D, Figure 15, shows that the looping period increased from about five days prior to 10 September to nearly 10 days during 10–30 September, then back to 5 days (this apparent change in periodicity could be the result of missing TOAs). While this time series shows that the float was looping continuously in the same eddy starting around 27 August 2006, we will focus on the eddy properties estimated with the wavelet ridge analysis only after the eddy properties stabilize, around 26 September and continuing to 18 December 2006 (solid green curves in Figure 14).

During this time R oscillated between 20 and 30 km, and V strengthened with time, reaching a maximum near −0.38 m s$^{-1}$ at R ~ 28 km. There is no linear relationship between R and V during this time (not shown). This combined with the somewhat higher temperature variance compared to when the eddy formed, Figure 14d (vertical blue lines), suggest that this float was near the radius of maximum azimuthal velocity, i.e., at the edge or just outside the solid body core.

Figure 16 shows two time sequences of ADT and float trajectories, the first seven panels for the formation of Eddy C, and the second set of five for the time when float looping in Eddy C stopped, with a break of about 2.5 months in between, during which Eddy C drifted slowly westward at speeds generally less than 0.05 m s$^{-1}$ (Figure 14e), as shown in Figure 13a. The trajectories of all three floats that interacted with Eddy C are shown here: 680 (dot), 582 (x) and 586 (*). All were within 20 dbar of each other before and during eddy translation.

Starting on 16 August 2006, Float 680 was drifting relatively rapidly (0.15–0.30 m s$^{-1}$; Figure 14e) southward over the continental slope toward the TGB just prior to being caught up in the formation of Eddy C. The other two floats were also drifting rapidly toward the TGB, 582 in the lead and 586 behind. The low-ADT trough east of the Grand Banks extended southward as the floats approached the TGB, 16 August – 6 September. As was the case when float 581 approached the TGB and Eddy B formed, there was an ADT gradient perpendicular to the slope at the TGB, indicating an offshore flow at least at the surface, 30 August 2006. Eddy B followed the direction of the offshore flow toward the southeast, whereas Eddy C and its floats drifted around the TGB and then westward, crossing under the offshore surface flow, which appeared to be weakening with time (see e.g., 13 September 2006). The floats converged where they separated from the slope, and 586 overtook both 680 and 582 and executed one loop around them, 13–20 September 2006. But 586 was not in the eddy core, evidenced by the fact that it drifted westward at the outskirts of the two looping floats, but without looping or cusping.

The subsequent westward drift of Eddy C does not appear correlated with any particular feature of the surface geostrophic current, until the end of December, when a large Gulf Stream meander or warm-core ring moved northward, intersecting the path of the eddy, 27 December 2006 and 3 January 2007. Floats 680 and 586 were the first to stop looping and accelerated southward along the path of the Gulf Stream, consistent with the notion that both were closer to the edge of Eddy C than 582 (see below).

Unfortunately, the tracks of floats 582 and 586 are too noisy to work well with the wavelet analysis method. The reader is again reminded that the sampling rate of these floats (one fix per day) is not ideal for accurately measuring the kinematic properties of eddies with periods of 4–6 days, especially when a few consecutive fixes are lost due to poor sound source signals. However, we can take advantage of float temperature to say something about where each float was relative to the eddy center, assuming that the coldest anomaly would be near the eddy center.

Figure 17 shows the temperature of the three floats as a function of time, starting when they were approaching the TGB on 16 August 2006. All three floats indicate similarly low temperature until about 10 September, when 586 made its one loop around the eddy and measured warmer temperatures. Float 586 was then temporarily left behind as Eddy C, with 582 and 680 looping in its core, translated rapidly westward. Temperatures of the three floats fluctuated around until the end of October, with 586 showing the largest temperature variance, consistent with its position at the periphery of the eddy. At the beginning of November, a steady pattern developed that persisted until looping stopped. Float 586 measured increasingly warmer temperatures, consistent with a radial position at the outskirts of the eddy. Its temperature increased progressively toward the mean temperature at 700 dbar, 5°C. Float 582 was consistently about 0.1°C colder than 680 starting at the beginning of November, suggesting that it was looping closer to the eddy center.

But both 582 and 680 temperatures were relatively constant (unlike 586) indicating little mixing with the ambient water. At the beginning of January 2007, floats 586 and 680 suddenly stopped looping (Figure 16), temperature shot up and the float accelerated southward along the Gulf Stream path, while float 582 continued looping in Eddy C for about two more weeks before it too stopped looping, accelerated southward and measured much higher temperatures. The large temperature fluctuations during the temperature transition may reflect rapid stirring of the eddy core water with the background as the eddy interacted with the Gulf Stream.

Summarizing the results described in this section, a few float trajectories obtained as part of a larger study of LSW spreading pathways in the depth range 700–1500 m have revealed for the first time the formation of energetic anticyclonic eddies at the TGB that transport cold water from the WBC of the subpolar gyre away from the continental slope and into the ocean interior. One such eddy, observed with a 1500-dbar float, crossed under the Gulf Stream, carrying a core of colder water into the subtropical interior. Mean float looping radii were in the range 13–26 km, and mean azimuthal speeds as high as 0.28 m s$^{-1}$ were observed. The three eddy formation events took place over a one-year time period. Comparisons of the eddy trajectories with maps of ADT from AVISO show that these eddies generally move in the same direction as the surface geostrophic current when the latter is well-defined.

## 3.2. Subpolar Water Eddies Observed in a Moored Array

### 3.2.1. Observations

In this section, we describe observations from the East and South sites of the BIO-50W moored array (Figure 1) that indicate the passage of at least two anticyclonic eddies with kinematic properties similar to the eddies described above. Contoured sections of the available moored temperature data vs. time and depth in Figure 18 provide an overview of the anomalies discussed in this paper. In January 1979 two distinct cold anomalies occurred at 1500 m at the East site separated by about 10 days; they extend in the vertical to 1200 m but not to 1000 m (Figure 18a). In July–August 1979 a 1000 m temperature anomaly with a cooling of more than 4°C appeared at the East site (Figure 18b) and penetrated to at least 1200 m depths (Figure 18b). Finally, during the same July–August 1979 period a weaker cold anomaly appeared at 1000m at the south site; there were no 1200 m data for this event (Figure 18c). Hendry (1981) briefly discussed some aspects of these events.

Quantitative analyses were carried out over the common periods for all instruments for a specific event at a specific mooring during which the anomalies at one or more depths stood out from the background conditions. Table 2a gives the analysis periods selected for the four cases outlined above: East 1a and East 1b refer to two January 1979 periods at the East site, East 2 refers to a 7.5-day period in late July and early August 1979 at the East site, and South 2 refers to a 3.25-day period at the South site. Background time series for temperature were defined by linear fits using representative end members on either side of the analysis periods. The selection of analysis periods and end members were necessarily somewhat subjective. Table 2b gives details of the background values and observed extreme temperature anomalies for nine cases: East 1a and East 1b anomalies at 1200 m and 1500 m, East 2 anomalies at 1000 m, 1200 m, and 1500 m, and South 2 anomalies at 1000 m and 1500 m.

### 3.2.2. θ–S properties of the anomalies

The conductivity data provide salinity and density time series when combined with measured temperature and measured or estimated pressure. No reliable information on the accuracy of the moored salinity was available so corrections were derived by comparing the moored salinities with spatially-interpolated values from the HydroBase 2 annual climatology (Curry, 2002). Moored salinity anomalies were calculated relative to site-specific reference HydroBase potential temperature (θ) – salinity (S) curves and an assumed bias equal to a robust estimate of the mean salinity anomaly was removed from each record. The resulting offsets defined independently for each instrument ranged from 0.06 to 0.27. As noted above, the 1200 m salinities from both East and South sites proved unusable because of noise in the conductivity data.

Background time series for salinity and potential density relative to 1500 dbar ($\sigma_{15}$) were defined over the same analysis periods and in the same way as for temperature. Potential temperature, salinity, and $\sigma_{15}$ values at the times of the temperature extremes are included in Table 2b. Both East 1a and East 1b anomalies showed a freshening of −0.2 at 1500 m at the times of the extreme temperature anomalies. Fresh anomalies of −0.9 and −0.4 at 1000 m were associated with the extreme temperature anomalies during the East 2 and South 2 events respectively. The 1500 m salinity anomalies during the East 2 and South 2 events were at the 0.01 noise level.

Figure 19a shows θ–S scatter plots for data from the combined 1500 m East 1a and 1b analysis periods and the 1000 m East 2 and South 2 analysis periods superimposed on the East site reference curve. In each case, the θ–S values associated with the anomalies show a nearly linear relationship that can be interpreted as a mixing line between two water types (Sverdrup et al., 1942). Least-squares fits of salinity vs. potential temperature gave R-squared values of 0.96 or more for each of the three cases. *CSS Hudson* 89037 CTD data at 10-dbar vertical resolution from January 1990 stations along 50°W from the Grand Banks southward to 38°N also shown in Figure 19 provide an example of the characteristic regional θ–S variability. The salient features are a warm, saline limb that closely follows the East site HydroBase reference curve and a cold, fresh limb whose coldest and freshest points are associated with the polar water carried by the near-surface Labrador Current. For the East 2 anomaly, the intersection of the mixing line with the reference θ–S curve defines a warm, saline end member with $\sigma_{15}$=34.31 kg m$^{-3}$. The depth of this $\sigma_{15}$ surface in the *Hudson* 89037 data set varies from 80 m near 42.5°N north of the Gulf Stream to a maximum of 1150 m near 39°N. A nominal cold, fresh end member is defined by averaging *Hudson* 89037 data points with salinity less than 35.5 that fall within 0.015 kg m$^{-3}$ of the mixing line. These data points come from 80–260 m depths at Stations 28–31 in water depths 550–2930 m and latitudes 42.8–42.3 N. Table 2c shows the end point values and estimated mixing rations for the 1000 m East 2 analysis and similar analyses carried out for the combined East 1a and East 1b anomalies and the South 2 anomaly. The individual data points do not in general fall exactly on the fitted mixing line and there is no unique way to project them onto the line, so mixing ratios based on both potential temperature and salinity are shown. The differences are not large. The potential temperatures on the South 2 mixing line are about 0.6 C warmer than values on the East 2 1000 m mixing line at same salinity in the 34–35 salinity range. The time evolution of the θ–S properties of the East 2 and South 2 anomalies is clearer in Figure 19b which shows the same data presented as a scatter plot of $\sigma_{15}$ vs. salinity anomaly relative to the reference curve. At both sites, the $\sigma_{15}$ values start out less than the mixing line value for the same salinity anomaly (equivalent to the same potential temperature), reach their extreme values, and then continue at values greater than the mixing line values.

### 3.2.3 Flow properties of the anomalies

To set the context for interpreting the flow anomalies, suppose a circularly symmetric eddy translated by a constant background flow moves past a mooring. A related kinematic model is discussed below. The sum of the background flow and the anomaly due to the eddy will be called the composite flow. Adopting a coordinate system with the x-axis oriented in the direction of eddy movement, the x-component of flow is called the longitudinal component (ut) in the discussion that follows. The flow in the y-direction of this right-handed (x, y) coordinate system is called the transverse component (vt). The composite longitudinal flow time series is symmetric about the time of closest approach and so attains an extreme value at that time except in the special case when the eddy passes directly over the observing site and the longitudinal flow is constant for all time. The corresponding transverse flow time series has a zero crossing at the time of closest approach, is anti-symmetric about that time, has zero mean value, and attains two extreme values at times equidistant from the time of closest approach with equal speeds but opposite signs.

The currents during the analyses periods associated with the temperature anomalies did in fact show multiple rate maxima consistent with this simple scheme. Table 2d lists the values and times of multiple rate maxima observed for seven of the nine cases in Table 2b that had valid rate measurements during the analysis periods. The associated flow directions at 1200 m and 1500 m at the East site changed by between −100° and 180° between successive rate maxima for the January 1979 events. Similar direction changes of −160° at 1000 m and −65° at 1500 m occurred at the East site during the July − August 1979 event. A smaller direction change of about 40° was noted at 1000 m at the South site for the later period. The elapsed times between the flow maxima give a quantitative characterization of the time scales of the flow disturbances. The maximum density signal associated with a passing eddy would be expected to occur at the time of closest approach, mid-way between the maxima in the extremes in transverse flow. In all seven cases the times associated with the extreme temperature events in Table 1b fall between the times of the corresponding rate maxima in Table 2d.

### 3.2.4 Eddy model

A simple kinematic model was applied to the flow fields to derive a parametric description of the best-fitting eddy for each analysis period. The approach is conceptually the same as White and McDonald's (2006) method for describing isolated eddies from time series measurements at a single current meter.

A circularly-symmetric Gaussian eddy model was chosen for the fitting procedure. The associated streamfunction has two free parameters, streamfunction amplitude A and spatial scale R.

$$\Psi(r) = A \exp(-\frac{1}{2}\left(\frac{r}{R}\right)^2)$$

A>0 gives a high-pressure core, anticyclonic eddy. For geostrophic flow, Ψ/f is the geopotential anomaly, where f is the Coriolis parameter. The associated geopotential height is (f/g)*Ψ, where g is the acceleration due to gravity.

The associated azimuthal flow vr is given by

$$vr(r) = -A\frac{r}{R^2}\exp(-\frac{1}{2}\left(\frac{r}{R}\right)^2) \; vr(r) = -A\frac{r}{R^2}\exp(-\frac{1}{2}\left(\frac{r}{R}\right)^2)$$

with the sign convention that vr is positive for cyclonic flow. The azimuthal flow reaches an extreme value V at r=R given by

$$V = -\frac{A}{R}\exp(-\frac{1}{2})$$

The speed of the azimuthal flow drops to 45% of its maximum value at 2R and to 5% of its maximum value at 3R. Any two of the parameters A, R, and V completely specify the eddy. Since R and V are related to directly measurable quantities, they are used in the fitting procedure described below.

We assume as discussed above that an eddy with fixed R and V moves past a mooring at a constant background velocity $\underline{U_b}$. Although the background flow field will in general also be evolving in time, the assumption is that the changes in background flow during the few days when a particular eddy dominates the flow field will be small enough for such a model to give interpretable results.

In addition to V and R, the composite flow model must also determine a best-fitting background flow speed $U_b$ and direction, the horizontal displacement $y_0$ of the eddy center from the mooring at its closest approach, and the corresponding time of closest approach $t_0$. The east and north flow components can then be specified as a function of time by a system of non-linear equations that are governed by these six free parameters. Since the fits are carried out in the time domain, R and $y_0$ appear only as the equivalent time intervals $R/U_b$ and $t_0/U_b$. The best-fitting parameters were determined for each case by a multidimensional unconstrained nonlinear minimization based on the Nelder–Mead simplex (direct search) method as implemented by the Matlab® procedure *fminsearch*. The cost function used was the sum of the squared differences between the observed and model u- and v-components during the defined fitting period. Starting values of R, V, $U_b$, background flow direction, $t_0$, and $y_0$ for the minimization were derived from initial exploratory analyses such as presented in Table 2d.

The fitting procedure converged to solutions representing between 93% and 97% of the composite flow for the seven cases treated in Table 2d. It failed to converge for a variety of starting values when applied to the East 1a and East 1b 1000 m cases which showed no sign of eddy activity. Table 3a lists the parameters and associated standard errors derived from the seven independent fits. Standard errors were estimated by a delete-one jackknife analysis carried out as part of the fitting procedure. The estimated standard errors had median values of about 10% of the corresponding optimum value for R and V. The largest of the standard errors for the times of closest approach (not shown) of 2 hours was obtained for the East 1a case at 1200 m. There was a small but distinct change in 1500-m flow direction during the South 2 event (no rate data were available) that mirrored the disturbance at 1000 m.

The 1000-m South 2 case stands out in that the estimated distance of closest approach is almost twice the estimated radius of maximum azimuthal flow, indicating a peripheral impact rather than a direct hit. The fitted radial scale is similar to the 1000-m East 2 value and since the two events were closely separated in time it seems possible that they represent the same feature. The background 1000-m flows during the peak the East 2 and South 2 anomalies were 0.12 m s$^{-1}$ towards 233°T and 0.20 m s$^{-1}$ towards 204°T respectively. The 3.2 day time lag between the times of closest approach gives a propagation speed of 16 km/day or 0.18 m s$^{-1}$ for the 50 km spatial separation, similar to the estimated background flow rates. The kinematic model gave a good fit to the 1000-m South 2 observations with maximum observed and modeled azimuthal flow anomalies of up to 0.35 m s$^{-1}$. However, the hypothetical eddy emerging from the fitting procedure has a maximum azimuthal flow anomaly with speeds greater that 0.6 m s$^{-1}$ between the eddy center and the South site. The Rossby number of more than 0.5 associated with this fit is unphysically high for a stable structure of that spatial scale. The estimated standard error of V for the 1000-m South 2 fit was notably larger than for any of the other fits (Table 3a). Although the fit was optimal in a least-squares sense, the suggestion is that it is sensitive to the assumption of circular symmetry and that an elliptical eddy model might give as good a fit without an artificially high value of V.

Table 3b gives values of the fitted streamfunction amplitudes expressed as geopotential height. Assuming a hydrostatic balance between the vertical pressure gradient and the buoyancy forces and a simple vertical structure, the pressure and density anomalies might be expected to have similar horizontal structures. This would also apply to temperature to the extent that it influences density. This motivated a regression of the temperature anomalies against a normalized version of the Gaussian streamfunction from the corresponding flow fit. The resulting model temperature anomalies at the eddy center and at the observing site are also given in Table 3b. The temperature fits accounted for more than 80% of the variance in the temperature anomaly time series during the analysis periods for six of the seven cases. The temperature fit for Case East 2 at 1500 m had residual variance compatible with the other fits and accounted for about 50% of the variance associated with a 0.1°C positive temperature anomaly.

In the East 2 analyses, the fitted streamfunction amplitude at 1500 m was about 0.27 of the 1000-m value. The more than two-fold change in the Brunt–Väisälä frequency N between 1000 and 1500 m confuses the analysis. The stretched vertical distance (Leaman and Sanford, 1975) between the 1000 m and 1500 m levels relative to the mean stratification is about 540 m. If we assume that the vertical structure of the eddy can be modeled with a Gaussian shape $\exp[-0.5*(z-z0/h0)]^2$ in the stretched coordinate system and that the 1000 m mooring sampled the center of the eddy so that z0 is known, the decrease in streamfunction requires h0=335 m in the stretched coordinate system. The Burger number $Bu = (NH/fL)^2$ (e.g. Pedlosky, 1987) is a dimensionless parameter related to the aspect ratio (H/L), where H and L are vertical and horizontal length scales. The two-point vertical fit gives $Bu^{1/2}$=0.69 using the estimated eddy radius at 1000 m as the horizontal scale. The calculated Burger number is independent of the reference stratification used to define the vertical stretching. For the earlier Events 1a and b the mean fitted streamfunction amplitude at 1200 m was about 0.47 of the 1500-m value and the stretched vertical distance between the 1200 m and 1500 m levels relative to the mean stratification was about 262 m. A similar calculation gives an h0 of 214 m in the stretched coordinate system and $Bu^{1/2}$=0.55 using the estimated eddy radius at 1500 m as the horizontal scale. The East 1a, b event had no measurable impact at 1000 m. The Gaussian model is reasonable consistent with this observation, giving a maximum azimuthal flow of less than 0.01 m s$^{-1}$ at 1000 m.

Figure 20 shows the complete results for the East 2 case including the estimated trajectory of the eddy center past the mooring site. Note that for fixed A and R, the eddy model can perfectly match any single flow vector with magnitude less than or equal to V with two different location of the eddy center relative to the location of the mooring. Both locations give the same bearing relative to the mooring but one is located at a distance r<R from the mooring and the other at a complementary distance r>R. For example, for R=10 km and V=−0.3 m s$^{-1}$, the azimuthal flows at r=5 km and 16.1 km are the same −0.08 m s$^{-1}$. The fitting procedure is insensitive to this ambiguity but a post-fit choice of the appropriate eddy centre at each time step is required to specify a unique eddy trajectory. The outer matching flow vector was selected on the approach and retreat of the eddy and the inner matching flow vector used on either side of the time of closest approach. There was little ambiguity in the appropriate choices except very close to the radius of maximum flow, in which case the flow vectors were nearly identical so there was little practical effect. An abbreviated Figure 21 shows results for only the transverse and longitudinal flow components and temperature for the South 2 case. Figure 22 shows companion vector plots of the modeled and residual background flow for the same two cases.

Table 3c lists derived parameters including the Rossby number defined as Ro=V/(f*R). As noted above, the South 2 anomaly stands out as having an unphysically high Rossby number. Values of N derived from the annual HydroBase climatology and scale heights (f/N)*R also appear in Table 3c. Since the radial scales are similar, the scale heights increase from 1000 m to 1500 m as the stratification decreases.

Table 3d returns to the salinity and density anomalies with the advantage that model streamfunction values and times of closest approach are now available. Observed $\sigma_{15}$ anomalies relative to background values were regressed against the normalized Gaussian streamfunction fits to give model $\sigma_{15}$ anomaly time series similar to the model temperature anomaly time series. Model salinity values could have been calculated in the same way but instead were computed as the salinities consistent with model temperature and $\sigma_{15}$ values created by combining background fields and model anomalies. The salinity fits account for a large percentage of the salinity variance in all cases except for the 1500 m East 2 case which showed little if any salinity anomaly. The model $\sigma_{15}$ time series account for a smaller fraction of the observed variability because of the compensating effects of temperature and salinity on density, especially for the 1000 m East 2 case where near-perfect compensation was achieved. Estimates of vertical displacement inferred from the difference between the instrument depth and the depth at which the observed or modeled $\sigma_{15}$ value occurred in the associated climatological HydroBase $\sigma_{15}$ profile were all less than 100 m except for the noise-dominated 1500 m East 2 case.

# 4. Discussion

## 4.1. Qualifying Subpolar Eddies as SCVs

Are the eddies observed in the RAFOS float and moored observations really SCVs? We can use the observations described above to formally answer this question, based on the criteria put forward by McWilliams (1985). He defined an SCV as having the following characteristics:

1. *A velocity structure with a localized subsurface maximum, as opposed to mesoscale eddies which are usually well-represented by the barotropic and first baroclinic modes.* The vertical structure of the eddies is illustrated by the Hovmöller diagrams in Figure 18. Even though the full vertical extent of the eddies is not resolved, Eddy 1a, b (Figure 18a), clearly has a subsurface velocity maximum.

2. *A horizontal scale that does not exceed the first baroclinic Rossby radius of deformation ($R_D$) and is sometimes smaller, whereas the radius of mesoscale eddies is usually larger than $R_D$.* From the mooring observations, the horizontal scale of the subpolar eddies, defined as the radius of maximum azimuthal velocity R, was estimated to be at most 15 km. From a global atlas of the first baroclinic Rossby radius of deformation (Chelton et al., 1998), $R_D$ is in the range 20–30 km between the TGB and 38°N, larger than the radial scale of the eddies. The Burger number can also be defined as the square of the ratio of the horizontal scale of the eddy to the deformation radius. That definition gives $Bu^{1/2}$ of order 0.5–0.75 for the observed eddies. The scale height is the least well-known property of the observed eddies due to limited vertical coverage. Our calculations above suggest $Bu^{1/2} \sim 0.5$ or 0.6 for the East 1a, b eddy with maximum observed amplitude at 1500 m and $Bu^{1/2} \sim 0.7$ for the East 2 eddy with maximum observed amplitude at 1000 m. McWilliams (1985) derived a $Bu^{1/2}$ of 0.9 for LDE Eddy D1.

3. *Axi-symmetric, anticyclonic circulation around a single maximum or minimum in geopotential anomaly (a monopole).* The float trajectories indicate mainly axi-symmetric circulation anticyclonically around an eddy center. The orbital motion may be elliptical in the presence of a background shear, although it may sometimes just appear to be elliptical due to the relatively low number of points defining each orbit.

4. *Maintain core properties over many rotation periods.* All three of the eddies observed with floats show evidence of property transport over long distances. Float 664 and Eddy A provide the most compelling example, with cold subpolar water transported 800 km south of the eddy formation site over about 30 rotation periods.

5. *Advected by combination of mean and mesoscale currents.* Comparison of the eddy trajectories derived from the float tracks with maps of ADT showed evidence of the eddies being advected by Gulf Stream meanders and mesoscale eddies. This was particularly evident with Eddy A. The arrival times of the East 2 and South 2 moored events at two sites separated by 50 km are consistent with a single feature advected by the background flow field.

## 4.2. Comparison with D1 from the Local Dynamics Experiment

The SCV D1 discussed in the Introduction was found at the southern edge of the recirculation near 31°N, 70°W (Elliott and Sanford, 1986a) centered at 1500 m with core salinity values of 34.975 and core potential temperatures relative to 1500 dbar near 4°C. The extreme minimum D1 salinity of 34.965 found between 1740 and 1840 m in profile G141 was spread over a range of potential temperatures relative to the surface of 3.838–3.917°C. The paper identifies the eddy center and the extreme salinity anomaly with the 34.65 kg m$^{-3}$ and 34.68 kg m$^{-3}$ $\sigma_{15}$ surfaces but Taft et al. (1986) note that the potential densities reported by LDE investigators were potential specific gravity anomalies based on the Knudsen (1899) equation of state which are systematically higher than the more-recent EOS-80 equation of state by 0.025 kg m$^{-3}$ because the two systems use different values for the maximum density of water. There are further differences between the Knudsen-Ekman and EOS-80 equations of state at higher pressures (UNESCO, 1991). The mean of the bounding potential temperatures 3.877°C for the extreme minimum salinity and the extreme minimum salinity give an EOS-80 $\sigma_{15}$ of 34.63 kg m$^{-3}$, about 0.05 kg m$^{-3}$ less than the reported value of 34.68 kg m$^{-3}$. The extreme salinity minimum corresponds to a salinity anomaly of -0.024 relative to the East site HydroBase reference θ-S curve.

The extreme East 1a, b 1500-m potential temperature 3.59°C and salinity 34.85 (Table 2b) were about 0.5°C cooler, 0.1 less saline, and 0.03 kg m$^{-3}$ lower in $\sigma_{15}$ than the corresponding 1500-m D1 core values. The East 1a, b anomaly was smaller and less energetic than Eddy D1; the model-fitted maximum azimuthal flows and associated radii of maximum azimuthal flows of 8±4 km and −0.14±0.03 m s$^{-1}$ respectively were both about 50% lower than Eddy D1. Eddy A sampled by 1500-m Float 664 had spatial and flow scales similar to Eddy D1, with a geometric-mean azimuthal flow of −0.29 m s[-1] and a geometric-mean radius of about 18 km. The estimated 3.63°C mean potential temperature in the core of Eddy A is close to the extreme value for the East 1a, b event anomaly and is also 0.5°C colder than Eddy D1 at 1500 m. The initial water properties of SCVs can vary greatly: the moored East 1a, b and East 2 − South 2 differences provide an immediate example. The initial flow disturbances and spatial scales would also be expected to have a range of values. Event East 1a, b and Eddy A could both represent the general class of events that might have created Eddy D1. If D1 had followed a path across the Gulf Stream from the north similar to Eddy A and translated directly from the vicinity of the seamounts to 31°N, 71°W (without hitting a seamount), a translation rate of 0.05–0.10 m s[-1] would cover the 2000 km involved over a time of order 1.1–1.7 years. Any deviation from a direct route would increase this age estimate, so it is not unreasonable to think of D1 as at least 1.5–2.0 years old.

Elliott and Sanford (1986a) suggested that the fresh anomalies observed in 1976 along 55°W on R/V Knorr Cruise 66 reported on by McCartney et al. (1980) were also possible analogues of Eddy D1. Knorr 66 55°W hydrographic data obtained from the World Ocean Data Base included 10 bottles at depths greater than 1200 m with salinity anomalies relative to the East site HydroBase reference of less than −0.03. These had median pressure 1507 dbar and median potential temperature, salinity, $\sigma_{15}$, and salinity anomaly of 3.76°C, 34.94, and 34.62 kg m[-3], and −0.043 respectively. The individual values cluster in θ–S space within or close to the triangle bounded by the East 1 a, b mixing line, the estimated Eddy A potential temperature, and the reference θ–S curve.

### *4.3. Subpolar Eddies Crossing the Gulf Stream*

Of the three eddies tracked by RAFOS floats, only one is tracked continuously across the Gulf Stream path and into the subtropical recirculation. This eddy was observed with a 1500-dbar float, whereas the other two eddies were tracked with 700-dbar floats. In one of these latter cases (Eddy C), the floats stopped looping when the eddy intersected the Gulf Stream. One possible explanation for this float behavior is that the eddy was destroyed by the horizontal and/or vertical shear of the Gulf Stream, at least at the level of the floats. If the eddy extends below the depths of strong shear, it's conceivable that the lower part could survive an encounter with the Gulf Stream and be swept downstream and possibly into the recirculation like Eddy A. Another possibility however is that the eddy is distorted by the local shear, but remains generally intact, and that the isobaric 700-dbar floats popped out of the top of the eddy as the eddy was subducted along the sloping isopycnals. This is illustrated schematically in Figure 23. As the eddy descends along the isopycnals, an isobaric float at 700 dbar maintains constant pressure and ends up above the eddy, no longer trapped in its anticyclonic rotation. Whether this happens or not will depend on the depth of the top of the eddy compared to the depth of the float. A float initially deeper in the eddy, such as float 664 at 1500 dbar in Eddy A, can remain trapped in the eddy as it crosses the Gulf Stream, although it will end up at a lower density (warmer temperature) within the eddy. As seen in Figures 4d and 7, this was what happened to float 664: at the same time Eddy A and float 664 intersected the Gulf Stream (mid-December 2005), float temperature increased by 0.2°C. Looping continued however, indicating that the float was still in the eddy.

## 4.4. Formation Process and Rate

Without more detailed observations of the three-dimensional flow field and density structure in the boundary current upstream of and around the TGB, it is not possible to determine the formation mechanism with any certainty. However, we present some additional information that may help to focus future attention to this area.

McWilliams (1985) proposed that the primary mechanism for generating SCVs was diapycnal mixing creating a volume of weakly-stratified fluid which was then injected into a more stratified background. Anticyclonic relative vorticity is then generated by geostrophic adjustment. However he also noted that instabilities of strong, narrow currents like the Labrador Current were a recognized source of eddies. Pickart et al. (1996) observed a lens-like eddy with a core of recently ventilated ULSW embedded in the Western Boundary Current near the northeastern corner of Flemish Cap. The eddy had a diameter of about 20 km, and only a very weak azimuthal velocity maximum ($0.005$ m s$^{-1}$). The eddy had been significantly eroded, evidenced by intrusions into the core. Using a primitive equation regional numerical model, Pickart et al. (1997) argued that such lenses could form from baroclinic instability of the baroclinic branch of the Labrador Current north of Flemish Cap and be swept downstream by the barotropic branch of the Labrador Current. In the model, this occurred when the baroclinic Labrador Current formed large meanders that eventually broke off anticyclonic eddies with radii 50–100 km. A subsurface velocity maximum developed as the mid-section of the eddy geostrophically adjusted to the higher offshore stratification. eddies were sheared by the advecting current to the point where their velocity structure was no longer coherent, but the patch of low-potential vorticity remained intact farther downstream. Pickart et al. (1996) predicted a relatively short lifetime for these eddies, on the order of a few months, based on an advective-diffusive model. This was seemingly inconsistent with much older ages based on chlorofluorocarbon measurements in the eddy core. This was explained in terms of an under-saturation of chlorofluorocarbons at the time of eddy formation.

The floats that ended up in eddies at the Tail of the Grand Banks did not show evidence of looping or cusping motion prior to separating from the slope. Thus they do not provide any evidence for a more northerly formation site for these eddies. However, the patches of low potential vorticity described in the Pickart et al. (1997) model study may contribute to eddy formation at the Tail of the Grand Banks. Considering the weak and eroded state of the observed lens of Pickart et al. (1996), the inconsistency in tracer age and modeled eddy lifetime, and the observations described in this paper, it seems possible that the Pickart lens could have been formed at the Tail of the Grand Banks and recirculated northward inshore of the North Atlantic Current.

It has been demonstrated theoretically and with laboratory models that along-stream variations in the steepness of the continental slope could cause instability of the boundary current and formation of eddies (Wolfe and Cenedese, 2006; Bracco et al., 2008). These ideas were applied to the west Greenland continental slope to explain the formation of anticyclonic Irminger Rings (Lilly et al., 2003). Figure 24 shows a close-up of the bathymetry along the eastern flank of the Grand Banks. The steepness of the slope (1000–3000 m) increases by about a factor of two from the northern to central region. The steeper slope extends for about 100 km, and then the slope broadens again at about 43.5°N. After about another 100 km in the along-isobath direction, the slope again steepens abruptly and turns more westward. In addition to these large-scale changes in steepness, the slope is indented by canyons and promontories which could also cause instabilities of the boundary current.

D'Asaro (1988) showed that a boundary current with negative relative vorticity that separates from the boundary at a sharp corner can spin up into an anticyclonic eddy with a relative vorticity comparable to that in the boundary current. The greater than 90° clockwise turn in the upper continental slope at the TGB (Figure 24) appears to make it a strong candidate for such separation of the WBC. Indeed the TGB was mentioned by D'Asaro as a possible site for the formation of SCVs via boundary current separation. D'Asaro (1988) demonstrated that the negative relative vorticity could be generated by frictional torque as the current flows along a boundary. Here we do not attempt to uncover the source of the negative vorticity, but simply compare its magnitude to that of the observed eddies. The vorticity of Eddy A was estimated to be $-2.8$ x $10^{-5}$ s$^{-1}$ based on the radial distribution of V (see section 3.1.1). Estimating the horizontal shear in the boundary current is more difficult. One of the most comprehensive studies of the boundary current system near the TGB was the analysis of the combined BIO and German moored arrays at 43°N (Schott et al., 2004; 2006). Unfortunately for the present study, these arrays were focused on the Deep WBC transport and its variability, and the instrumentation did not extend far enough up the continental slope or up into the water column to resolve the horizontal shear of the flows along the upper continental slope. Fratantoni and Pickart (2007) developed a comprehensive climatology of the velocity and hydrographic structure of the shelf-break currents from Greenland to the Mid-Atlantic Bight, but her analysis extended only to 700-m depth and by necessity considered only the baroclinic flow components.

One could imagine using the *ExPath* floats themselves to estimate the horizontal structure of the boundary current where it approaches the TGB (see e.g., Fischer and Schott, 2002), but there were only 15 700-dbar and five 1500-dbar floats that drifted southward over the eastern slope of the Grand Banks, and the velocity observations are noisy and lacking in a well-defined horizontal structure (not shown). We can only crudely estimate the lateral shear in the boundary current by using a representative speed of floats upstream of the TGB, before they were caught in eddies. Using a speed of about 0.20 m s$^{-1}$ (Figures 4e and 14e), as representative of the jet maximum, and assuming the velocity goes to zero at the boundary over a distance of ~10–20 km (Figure 25), gives a relative vorticity on the inshore flank of the current of −1 to −2 x 10$^{-5}$ s$^{-1}$, which is of the same order as the estimated relative vorticity of the eddy cores. With this crude estimate we can only argue that it seems possible that there is sufficient shear in the boundary current to produce eddies with the observed rotation periods. Clearly more observations of the boundary current at this critical location are needed to better understand the eddy formation process.

Finally, the observations presented in this paper raise the question of how important these SCVs are in the ventilation of the interior ocean and southward transport of subpolar waters. The answer will of course depend on the rate at which these eddies form and the volume of the anomalous core water, neither of which can be accurately estimated with the present observations. There is reason to believe eddy formation may occur relatively frequently. Three eddies were observed to form at the TGB over a 12-month period with only a dozen or so floats passing through the formation region. Two eddies passed by the BIO-50W moored array in less than one year. Using a time scale of one month for an eddy to form and pull away from the TGB (see e.g. Figure 7) and assuming only one eddy can form at a time, we estimate an upper bound of about 12 eddies formed per year. Considering that these eddies are capable of trapping a core of subpolar water for perhaps at least two years, there could be a standing crop of about 24 eddies. A companion study is underway to use historical CTD profiles from ship surveys and Argo floats to examine the distribution and population of these SCVs in the subtropical western North Atlantic, as Richardson et al (1991) did for meddies in the eastern Atlantic.

## 5. Summary

Three anticyclonic eddies were observed to form at the Tail of the Grand Banks between October 2005 and September 2006 with eddy-resolving RAFOS float trajectories at 700 and 1500 dbar. The first eddy (A) was tracked with a 1500-dbar float as it translated southward from the Tail of the Grand Banks, under the Gulf Stream and into the subtropical recirculation. Mean looping radius, speed and period were 8.3 km, $-0.29$ m s$^{-1}$ and 5.3 days, respectively. The eddy core was in solid body rotation and carried anomalously cold water (temperature anomaly of $-0.5$ to $-0.7°C$) into the subtropics. The vorticity Rossby number for the eddy was estimated to be 0.3. The eddy hit one of the Corner Rise Seamounts near 35°N, which caused the float to be lost from the eddy.

The other two eddy formation events (B and C) were observed with 700-dbar floats. Eddy B was only tracked for one month as it translated rapidly southeastward from the Tail of the Grand Banks along the flank of the Southeast Newfoundland Ridge. Eddy C was observed with three 700-dbar floats, two of which were trapped in the eddy core. Unlike the previous two eddies, Eddy C translated westward from the TGB, staying north of the Gulf Stream for four months, until it hit a northward meander of the Gulf Stream. All float looping stopped at this point, either because the eddy was sheared apart (at least at 700 dbar) or the floats popped out of the top of the eddy as it descended along sloping isopycnals. Mean kinematic properties of Eddy C were radius 26 km, speed −0.32 m s⁻¹, and period 6.7 days.

At least two distinct cold-core anticyclonic lenses with similar kinematic properties passed through a small moored array deployed late in 1978 south of the Gulf Stream near 38N, 50W. The lenses were identified by anomalies in temperature, salinity, and horizontal flow at moored instruments located at depths between 1000 m and 1500 m. The first lens, which passed close to the easternmost mooring in January 1979, had a radial scale of about 10 km and maximum eddy speeds of order 0.15 m s⁻¹. The largest anomalies were at 1500 m, with a slightly reduced influence at 1200 m and no measurable effects at 1000 m. Anomalies observed in late July – early August 1979 at two moorings separated by 50 km can plausibly be attributed to the passage of a single lens with radial scale 10–15 km and maximum speeds of 0.25–0.35 m s⁻¹. The maximum impact was at 1000 m, with reduced but noticeable effects at both 1200 m and 1500 m. This lens appeared to pass directly over the easternmost site and then continued to the southwest with the prevailing background flow to glance the southernmost site about three days later. A mixing analysis suggests that the core temperature and salinity properties of the 1000-m anomalies contained as much as 85% pure Labrador Current Water such as found in the upper 250 m at the shelf break or in the eastward recirculation of Labrador Current Water that occurs south of the TGB.

The observations described in this paper confirm earlier speculation that the Tail of the Grand Banks is a formation site for small, coherent, long-lived eddies containing a core of relatively cold, fresh subpolar water. One float trajectory reveals for the first time how such SCVs can transport the subpolar water under the Gulf Stream and into the subtropics. The θ–S properties of eddy core waters as seen at moored instruments south of the Gulf Stream can be traced back to the Western Boundary Current in the vicinity of the Tail of the Grand Banks. The small radial scale of the observed eddies indicates that they are part of the class of eddies known as submesoscale coherent vortices (SCV). The kinematic properties of the eddies are similar to an SCV, called D1, with a core of cold, fresh water studied intensively during the Local Dynamics Experiment near 31°N, 70°W and are used to estimate an age for D1 of about two years. The leading hypothesis for the formation of the SCVs at the Tail of the Grand Banks is that the Western Boundary Current separates from the continental slope where the isobaths turn sharply westward, and wraps up into an eddy due to the negative vorticity on the inshore flank of the Western Boundary Current. The fortuitous observations described here are not sufficient to determine with any certainty the cause of eddy formation or the potential importance of these eddies to the ventilation of the interior subtropical North Atlantic with any confidence. A dedicated study that includes observations of the time-dependent velocity and density structure of the middle and upper continental slope near the Tail of the Grand Banks is needed.

**Acknowledgements**

The authors gratefully acknowledge H. Furey for her expert tracking of the RAFOS floats used in this study and for the preparation of figures and proof-reading. This work was funded by grant numbers OCE-0824652 to A. Bower at WHOI and OCE-0751697 to J. M. Lilly at NorthWest Research Associates by the U.S. National Science Foundation. The altimeter products were produced by Ssalto/Duacs and distributed by Aviso with support from CNES. The co-authors would also like to express special gratitude to the contributions of Dr. H. Thomas Rossby to this work. Tom's pioneering work in the development of acoustically tracked subsurface floats is the foundation on which all Lagrangian studies of large- and meso-scale ocean circulation rest. With the high-resolution trajectories afforded by acoustic tracking, Tom and others have uncovered a rich variety of phenomena that have fundamentally changed how we think about ocean circulation. With his spirit of generosity and dedication to the success of others, Tom Rossby is a mentor and colleague that we hold in the highest regard.

# Tables

Table 1a.  Dates of eddy observations with RAFOS floats.

| Event | Eddy Detected | | Eddy Properties Estimated | |
|---|---|---|---|---|
| | Start | End | Start | End |
| Eddy A (664) | 22 Oct 2005 | 30 Mar 2006 | 02 Dec 2005 | 18 Mar 2006 |
| Eddy B (581) | 22 Jun 2006 | 24 Jul 2006 | 03 Jul 2006 | 12 Jul 2006 |
| Eddy C (680) | 27 Aug 2006 | 28 Dec 2006 | 26 Sept 2006 | 18 Dec 2006 |
| Eddy C (582) | 13 Aug 2006 | 04 Jan 2007 | – | – |

Table 1b.  Eddy properties estimated from wavelet analysis for the time period of reliable statistics, except for mean temperature, which is estimated for the time period of eddy detection (see Table 1a).

| Event | Mean R (km) | Mean V ( m s$^{-1}$) | Mean T (days) | Mean *In situ* (potential) Temp. (°C) | Lifetime (months) |
|---|---|---|---|---|---|
| Eddy A (664) | 18.3 | −0.29 | 5.3 | 3.75 (3.63) | 5.1 |
| Eddy B (581) | 13.4 | −0.16 | 7.3 | 4.04 (3.99) | 1.1 |
| Eddy C (680) | 25. | −0.3 | 6. | 3.87 (3.82) | 4. |
| Eddy C (582) | − | − | − | − | 4.8 |

Table 2a.  Fitting periods.

| Case | Mean depths | Fit period start | Fit period start | Fit period duration | Number of points in time window |
|------|-------------|------------------|------------------|---------------------|--------------------------------|
|      | m | UTC | UTC | days | |
| East 1a | 1021 1233 1517 | 18:00 30 Dec 1978 | 12:00 04 Jan 1979 | 4.75 | 20 |
| East 1b | 1021 1233 1517 | 00:00 11 Jan 1979 | 00:00 13 Jan 1979 | 2.00 | 9 |
| East 2 | 1021 1233 1517 | 18:00 25 Jul 1979 | 06:00 02 Aug 1979 | 7.50 | 31 |
| South 2 | 1017 1513 | 18:00 30 Jul 1979 | 00:00 03 Aug 1979 | 3.25 | 14 |

Table 2b.  Observed extreme temperatures and associated fields.

| Case | Depth | Temperature | | | | Extreme values | | |
|---|---|---|---|---|---|---|---|---|
| | | Extreme | Reference | Anomaly | Time of extreme | $\theta$ | Salinity | $\sigma_{15}$ |
| | m | °C | °C | °C | UTC | °C | | kg m$^{-3}$ |
| East 1a | 1233 | 4.69 | 6.05 | -1.36 | 1800 01 Jan 1979 | 4.59 | – | – |
| East 1a | 1517 | 3.71 | 4.63 | -0.92 | 1800 01 Jan 1979 | 3.59 | 34.86 | 34.59 |
| East 1b | 1233 | 3.90 | 5.88 | -1.98 | 0000 12 Jan 1979 | 3.80 | – | – |
| East 1b | 1517 | 3.71 | 4.61 | -0.91 | 1800 11 Jan 1979 | 3.59 | 34.85 | 34.58 |
| East 2 | 1021 | 2.37 | 6.75 | -4.38 | 0000 29 Jul 1979 | 2.31 | 34.15 | 34.20 |
| East 2 | 1233 | 4.47 | 5.38 | -0.90 | 0000 29 Jul 1979 | 4.37 | 34.77 | 34.41 |
| South 2 | 1017 | 4.78 | 6.85 | -2.07 | 0600 01 Aug 1979 | 4.70 | 34.56 | 34.20 |

Table 2c.  θ–S mixing analysis.

| Case | Depth | Warm end saline point | | | Cold and fresh end point | | | | Cold fraction | |
|------|-------|------|------|-------------|------|------|-------------|----------|------|------|
| | | θ | S | $\sigma_{15}$ | θ | S | $\sigma_{15}$ | pressure | θ | S |
| | m | °C | | kg m$^{-3}$ | °C | | kg m$^{-3}$ | dbar | | |
| East 1a, b | 1517 | 4.60 | 35.03 | 34.58 | 3.37 | 34.81 | 34.58 | 738 | 82% | 88% |
| East 2 | 1021 | 6.63 | 35.10 | 34.31 | 0.91 | 33.85 | 34.13 | 175 | 76% | 76% |
| South 2 | 1017 | 7.26 | 35.13 | 34.22 | 0.13 | 33.48 | 33.91 | 135 | 36% | 35% |

Table 2d.  Observed flow extremes.

| Case | Depth | First rate maximum | | | Second rate maximum | | | Elapsed time | Direction change |
|---|---|---|---|---|---|---|---|---|---|
| | | Rate | Direction | Time | Rate | Direction | Time | | |
| | m | m s$^{-1}$ | °T | UTC | m s$^{-1}$ | °T | | days | °T |
| East 1a | 1233 | 0.16 | 41 | 0000 31 Dec 1978 | 0.11 | 255 | 0600 04 Jan 1979 | 4.25 | −146 |
| East 1a | 1517 | 0.20 | 68 | 0000 31 Dec 1978 | 0.20 | 248 | 0600 04 Jan 1979 | 4.25 | −180 |
| East 1b | 1233 | 0.18 | 351 | 1200 11 Jan 1979 | 0.11 | 253 | 1800 12 Jan 1979 | 1.25 | −98 |
| East 1b | 1517 | 0.18 | 357 | 1200 11 Jan 1979 | 0.17 | 225 | 1800 12 Jan 1979 | 1.25 | −132 |
| East 2 | 1021 | 0.29 | 282 | 1200 27 Jul 1979 | 0.25 | 121 | 1800 30 Jul 1979 | 3.25 | −161 |
| East 2 | 1517 | 0.15 | 262 | 0000 27 Jul 1979 | 0.06 | 197 | 1200 31 Jul 1979 | 4.50 | −65 |
| South 2 | 1017 | 0.21 | 21 | 0000 01 Aug 1979 | 0.12 | 60 | 1200 01 Aug 1979 | 0.50 | 39 |

Table 3a.  Model parameters and standard errors.

| Case | Depth | Maximum azimuthal flow V | Radial scale R | Closest approach y0 | Time of closest approach | Background flow speed | Background flow direction |
|---|---|---|---|---|---|---|---|
| | m | m s$^{-1}$ | km | km | UTC | m s$^{-1}$ | °T |
| East 1a | 1233 | −0.12 ± 0.01 | 6.6 ± 0.7 | 2.5 ± 1.2 | 0000 02 Jan 1979 | 0.05 ± 0.03 | 337 ± 004 |
| East 1a | 1517 | −0.18 ± 0.02 | 13.7 ± 1.7 | 4.6 ± 3.1 | 2000 01 Jan 1979 | 0.07 ± 0.06 | 336 ± 019 |
| East 1b | 1233 | −0.11 ± 0.03 | 6.8 ± 1.7 | 1.6 ± 1.3 | 0000 12 Jan 1979 | 0.14 ± 0.02 | 292 ± 006 |
| East 1b | 1517 | −0.14 ± 0.02 | 5.8 ± 1.2 | 1.0 ± 1.2 | 0100 12 Jan 1979 | 0.10 ± 0.03 | 289 ± 004 |
| East 2 | 1021 | −0.24 ± 0.01 | 12.2 ± 0.5 | 3.1 ± 0.4 | 2300 28 Jul 1979 | 0.10 ± 0.01 | 205 ± 002 |
| East 2 | 1517 | −0.05 ± 0.00 | 14.7 ± 0.9 | 1.9 ± 2.7 | 0600 29 Jul 1979 | 0.09 ± 0.01 | 229 ± 001 |
| South 2 | 1017 | −0.61 ± 0.16 | 10.3 ± 0.9 | 18.5 ± 2.8 | 0700 01 Aug 1979 | 0.16 ± 0.03 | 224 ± 008 |

Table 3b.  Model streamfunction and temperature fits and goodness of fit statistics.

| Case | Depth | Geopotential height anomaly | Temperature anomaly | | Flow | | Temperature | |
|------|-------|------|------|------|------|------|------|------|
| | | Model amplitude | Model amplitude | At closest approach | rms residual | R-squared | rms residual | R-squared |
| | m | m | °C | °C | m s$^{-1}$ | | °C | |
| East 1a | 1233 | 0.02 | −1.0 | −1.0 | 0.03 | 0.93 | 0.27 | 0.88 |
| East 1a | 1517 | 0.05 | −1.0 | −0.9 | 0.03 | 0.96 | 0.15 | 0.97 |
| East 1b | 1233 | 0.01 | −2.2 | −2.1 | 0.02 | 0.97 | 0.11 | 0.99 |
| East 1b | 1517 | 0.02 | −1.0 | −1.0 | 0.02 | 0.97 | 0.10 | 0.98 |
| East 2 | 1021 | 0.06 | −5.1 | −5.0 | 0.04 | 0.95 | 0.58 | 0.96 |
| East 2 | 1517 | 0.02 | 0.1 | 0.1 | 0.02 | 0.95 | 0.08 | 0.49 |
| South 2 | 1017 | 0.12 | −6.6 | −1.3 | 0.02 | 0.97 | 0.39 | 0.81 |

Table 3c.  Derived parameters (Rossby number Ro, scale height h0).

| Case | Depth | Ro | Brunt–Väisälä period | N/f | h0 |
|------|-------|-----|------|-----|-----|
| | m | | h | | m |
| East 1a | 1233 | 0.16 | 0.73 | 27 | 246 |
| East 1a | 1517 | 0.12 | 1.17 | 17 | 825 |
| East 1b | 1233 | 0.14 | 0.73 | 27 | 254 |
| East 1b | 1517 | 0.21 | 1.17 | 17 | 349 |
| East 2 | 1021 | 0.17 | 0.49 | 40 | 309 |
| East 2 | 1517 | 0.03 | 1.17 | 17 | 886 |
| South 2 | 1017 | 0.51 | 0.47 | 41 | 252 |

Table 3d.  Observed and modeled salinity and density anomalies at closest approach.

| Case | Depth | Salinity anomaly at closest approach | | Salinity $R^2$ | $\sigma_{15}$ anomaly at closest approach | | $\sigma_{15}$ $R^2$ | Vertical displacement inferred from $\sigma_{15}$ anomaly | |
|---|---|---|---|---|---|---|---|---|---|
| | | Observed | Model | | Observed | Model | | Observed | |
| | (m) | | | | kg m$^{-3}$ | kg m$^{-3}$ | | (m) | (m) |
| East 1a | 1517 | −0.18 | −0.18 | 0.97 | −0.007 | −0.014 | 0.78 | −29 | −56 |
| East 1b | 1517 | −0.18 | −0.20 | 0.99 | −0.016 | −0.023 | 0.73 | −60 | −85 |
| East 2 | 1021 | −0.87 | −0.93 | 0.98 | −0.043 | −0.016 | 0.03 | −29 | −12 |
| East 2 | 1517 | 0.00 | 0.01 | 0.15 | −0.028 | −0.012 | 0.40 | −104 | −47 |
| South 2 | 1017 | −0.45 | −0.30 | 0.87 | −0.017 | −0.019 | 0.45 | −12 | −13 |
| South 2 | 1513 | −0.01 | | | −0.017 | | | −63 | |

# Appendix: Extracting Eddy Signals from Float Trajectories

## A.1. Fundamentals

It is convenient to let the float trajectory be expressed as a complex-valued time series

$$z(t) \equiv x(t) + iy(t)$$

where x(t) and y(t) are displacements, in kilometers, obtained from an expansion about a central latitude/longitude point in the usual way, and $i = \sqrt{-1}$ . This time series is assumed to be of the form

$$z(t) \equiv z_o(t) + z_e(t)$$

where $z_e$(t) is the oscillatory displacement of a particle about the center of an eddy, and $z_o$(t) is a residual corresponding to the apparent center of the eddy. In this representation, any measurement noise would be included in the residual signal $z_o$(t).

The key step is the introduction of a model for the eddy signal, which is represented as a particle orbiting the periphery of a time-varying ellipse (Lilly and Gascard, 2006; Lilly and Olhede, 2010):

$$z_e(t) \equiv e^{i\theta(t)}[a(t)\cos\phi(t) + ib(t)\sin[\phi(t)]]. \quad (1)$$

Here a(t) and b(t) are the time-varying major and minor semi-axis lengths, $\vartheta$(t) is the time-varying ellipse orientation, and $\phi$(t) is the phase angle of the particle location around the ellipse periphery. The semi-major axis a(t) is positive by definition, while the sign of b(t) determines the sense (counterclockwise or clockwise) in which the particle orbits the ellipse.

While the model (1) is under-determined, unique choices for the right-hand-side parameters can be found for a given left-hand side by employing the same logic by which a musical note (for example) may be described as having a time-varying frequency and amplitude; see Lilly and Olhede (2010) for details. The model has two innovations: it enables the eddy shape to be elliptical rather than circular, and more importantly it captures the possibility of time-dependency of the eddy currents. It is evident that the model (1), while purely kinematic, is an attractive match for a variety of dynamical solutions, for example, an anticyclonic lens deformed into an ellipse by a steady strain (Ruddick, 1987).

## A.2. Physical quantities

The ellipse properties can be transformed into more physically meaningful quantities. The geometric mean radius, $R(t) \equiv \sqrt{a(t)b(t)}$, is an average radius that does not change as the particle orbits a fixed ellipse; rather, it changes only as the ellipse geometry changes. Note that $R(t)$ is not intended to be an estimate of the radius of an eddy core. Rather, $R(t)$ can be interpreted as the apparent radius of the eddy streamline that is instantaneously occupied by the float. $R(t)$ can change either because the eddy evolves, or because the float moves to a new streamline within an unchanging eddy.

An analogous mean speed, the geometric mean velocity $V(t)$, similarly characterizes the typical speed with which the particle circulates the ellipse periphery. This quantity is found by explicitly differentiating (1) and rearranging the result to have the same form, but with a new pair of semi-axes $\tilde{a}(t)$ and $\tilde{b}(t)$, which now have units of velocity. Then $V(t) = \text{sgn}(\tilde{b}(t))\sqrt{\tilde{a}(t)\tilde{b}(t)}$ gives the geometric mean velocity, defined to be negative for clockwise rotation; exact expressions for $\tilde{a}(t)$ and $\tilde{b}(t)$ in terms of the parameters of (1) may be found in Appendix E of Lilly and Gascard (2006). The geometric mean velocity $V(t)$ agrees with the azimuthal velocity when the ellipse is purely circular, but is more appropriate for describing the velocity associated with an elliptical displacement signal.

A measure of the shape of the ellipse is given by

$$\xi(t) = \frac{2\, a(t)b(t)}{a^2(t) + b^2(t)}$$

which is naturally called the ellipse circularity, as $\xi(t)$ varies between $-1$ for a negatively rotating circle and $+1$ for a positively rotating circle. Purely linear motion corresponds to $\xi(t)\quad 0$.

A characteristic time-varying frequency associated with the motion of a particle around an ellipse can now be constructed. Lilly and Olhede (2010) show that

$$\omega(t) = \frac{d}{dt}\phi(t) + \xi(t)\frac{d}{dt}\theta(t)$$

is the natural measure of the instantaneous frequency content of the modulated ellipse $z_\bullet(t)$. This involves contributions from both the changing position of the particle around the ellipse $\phi(t)$, as well as the changing ellipse orientation $\theta(t)$. When the ellipse is purely circular with a fixed radius $R$, $\omega(t)$ becomes the (constant) angular velocity $|V|/R$. For a purely linear signal, as when $y(t)$ vanishes for example, $\omega(t)$ reduces to the standard definition of the instantaneous frequency content of a univariate signal. Weighted by the local power of the signal $z_\bullet(t)$, the time-average of $\omega(t)$ recovers the first moment (that is, the mean value) of the Fourier spectrum of $z_\bullet(t)$ – an attractive result that supports the interpretation of $\omega(t)$ as an "instantaneous" frequency. See Lilly and Olhede (2010) for details.

## A.3. Parameter estimation

In practice, the ellipse parameters appearing in (1) must be estimated from an observed trajectory $z(t) \equiv z_\circ(t) + z_\bullet(t)$ containing both the eddy signal as well as the background flow. A solution to this problem has been proposed by Lilly and Olhede (2010, 2009b) building on the earlier work of Lilly and Gascard (2006). The ellipse parameters are identified through an optimization problem in which the signal is projected onto a time- and frequency-localized oscillatory "test function", that is, a wavelet $\psi(t)$. The wavelet $\psi(t)$, which like a complex exponential $e^{i\omega t}$ is naturally complex-valued, can be rescaled in time as $\psi_s(t) \equiv (\psi(t/s))/s$ to generate different frequency content. At each moment, the particular choice of scale $s$ is found which maximizes the power of the projection. Chaining together the points of maximum projection from moment to moment leads to a time-scale curve known as a "ridge". The values obtained by the wavelet transform along the ridge form an estimate of $z_\bullet(t)$, from which estimated ellipse parameters can be deduced.

The estimation of $z_\bullet(t)$ from the trajectory $z(t)$ has only a handful of free parameters. One must specify a frequency band – essentially, a range of $\omega(t)$ values – to search for a local optimum, which we take to be from 1/2 to 1/16 of the local Coriolis frequency. There is an amplitude cutoff to reject a point from consideration for a local optimum if its value is too weak, which is chosen here as a one kilometer displacement signal; this cutoff has the effect of removing a few "false positives" which appear to be obviously due to noise.

The remaining, and most important, free parameter concerns the choice of wavelet $\psi(t)$. An investigation into different families of wavelets (Lilly and Olhede, 2009a) showed a particular family, dubbed the "Airy wavelets", to have particularly attractive properties for this type of analysis on account of their high degree of frequency-domain symmetry. The Airy wavelets are controlled by a single parameter, $P_\psi$, which is called the wavelet *duration*. As the duration increases, more "wiggles" fit into the central time window of the wavelet, and the degree of time-domain smoothing increases. Here, we choose $P_\psi = 3$, which corresponds to about one oscillation fitting between the half-power points flanking the central maximum of the wavelet amplitude. This wavelet can be seen in the last row of the "Airy wavelet" column of Figure 6 in Lilly and Olhede (2009a).

It may be shown that if one creates a displacement signal $z_e(t)$ through equation (1) given a prescribed set of ellipse parameters, the method described above can recover these parameters to a high level of accuracy provided the ellipse properties change slowly compared with the ellipse frequency (Lilly and Olhede, 2011), and provided the "noise" $\varepsilon_0(t)$ is not too strong in the frequency range occupied by the eddy signal at each moment.

## A.4. Analysis example

As an example, Figure A1 shows the recovery of a synthetic "eddy". A purely circular eddy having a constant period $2\pi/|\omega|$ of five days begins abruptly at time t=0 days. The eddy initially has radius R(0)=10 km and velocity V(0)=−0.14 m s$^{-1}$. Radius and velocity magnitude linearly increase, so that at time t=50 days both have doubled to R(50)=20 km and V(50)=−0.29 m s$^{-1}$. Signals such as this one are commonly observed to occur as a float slowly drifts across material surfaces in the solid-body core of an eddy.

The recovered properties, using identical parameter settings to those applied to the data, are shown in each panel with gray lines. There is an "edge effect" region, occurring within plus or minus one period from the sudden transitions at the start and end of the eddy signal. In these time intervals, the eddy properties are not accurately recovered. Away from these regions, the eddy properties are recovered extremely well. In Figure A1a, for example, the recovered displacement signals are not visible during most of the record because they are exactly overlapped by the true signals. Although in this example, the mean flow has been set to zero, the method is entirely invariant to the addition of a constant mean flow, i.e. a signal of the form $z_0(t) = u_0(t-t_0) + iv_0(t-t_0)$, where $u_0$ and $v_0$ are constants. Therefore this example suffices to illustrate the recovery of slowly-varying eddy properties in an arbitrary constant advecting flow.

The corresponding radius / velocity plot is shown in Figure A2, using the geometric mean radius and velocity defined above. There are two messages to this figure. The first is that slow changes of the radius and velocity can be accurately recovered, as is seen by the fact that the recovered properties (solid gray) are nearly completely overlapped by the true properties (black). The second message is that the edge-effect regions can lead to straight lines on the R/V plane. These regions give the *appearance* that the float is profiling through a solid body core, but this is an artifact of the analysis. Loosely speaking, one may say that in the edge effect regions the analysis "sees" a signal with decaying amplitude but fixed frequency. In summary, points near the edge effect regions are interesting in that the analysis method detects an oscillation, even though the properties of this oscillation cannot be accurately estimated because the local rate of change is too great.

# Figure Captions

Figure 1: Schematic diagram of the major currents around the Grand Banks of Newfoundland, including the southward-flowing Western Boundary Current of the subpolar gyre (blue) and the Gulf Stream (GS), its Northern and Subtropical Recirculations and the North Atlantic Current (red). *ExPath* RAFOS floats were released sequentially at several cross-slope positions along the section north of the Grand Banks near 50°N. The triangle near 38°N, 50°W indicates the position of the BIO-50W moored array. The location of one of the sound sources used for tracking (D) is also shown. Isobaths are at 1000-m intervals.

Figure 2: (a) Trajectory of ExPath RAFOS float 664 (Eddy A) after its release in the southward-flowing WBC near 50°N. The float trajectory is color-coded according to float temperature. The trajectory is superimposed on the climatological mean *in situ* temperature at 1500 meters from HydroBase, which is plotted with the same color scale as the float temperature on the trajectory. White dots are plotted at the first of each month. (b) Time series of float temperature and climatological mean temperature along the float's trajectory.

Figure 3: Residual trajectory (black curve) for float 664 and eddy realizations (ellipses, drawn to scale) every three days along the trajectory obtained from the wavelet ridge analysis. Ellipses from the "edges" of the eddy segments are black, indicating the lower confidence in the eddy dimensions within two rotation periods of either end of the eddy segments. Isobaths are drawn every 1000 m.

Figure 4:  Time series of Eddy A properties based on data from float 664 and the wavelet ridge analysis. The different colored lines correspond to the different eddy segments identified by the analysis. The property estimates are less reliable during the time it takes for two rotation periods at the beginning and end of each eddy segment, and these estimates are denoted by dotted lines. (a) geometric mean radius, R, (b) geometric mean speed, V, (c) rotation period, T=2π/ω, where ω is the angular velocity, (d) float temperature and (e) residual speed. The vertical lines in (d, e) bracket the full length of each eddy segment (including the first and last two rotation periods).

Figure 5:  Time-of-arrival (TOA) for signal from sound source D (see Figure 1 for location) as recorded by float 664. The continuous oscillatory signal indicates that float 664 was looping continuously in a single eddy throughout the period of low-quality trajectory (November 2005).

Figure 6:  V versus R from the analysis of float 664. Open circles indicate the ends of the eddy segments. The linear fit is made using only the solid green symbols.

Figure 7:  Time sequence of maps of  m s Dynamic Topography (ADT) from AVISO (combination of time-dependent sea level anomaly and mean  m s dynamic topography) with 14-day track segments of float 664 superimposed. Contour interval is 10 cm and isobaths are drawn every 500 m, down to 4500 m. Panels are separated by two weeks.

Figure 8: Expanded view of ADT (contour interval 5 cm) and float 664 track segments (14-day) as the float approached the Corner Rise Seamounts. Bathymetry is contoured at 500-m intervals down to 4500 m.

Figure 9: Silhouette of the Corner Rise Seamounts as viewed from the north, with the location of Eddy A when it hit one of the seamounts. Depths based on ETOPO2 gridded digital bathymetry.

Figure 10: (a) Same as Figure 2a but for the 700-dbar float 581 (Eddy B), superimposed on the climatological mean temperature at 700 m from HydroBase. (b) temperature measured by float 581 and mean climatological temperature at 700 m along the float path.

Figure 11: Same as Figure 3 but for float 581, Eddy B. Isobaths are contoured every 500 m.

Figure 12: Same as Figure 7 but for the time period of Eddy B. Panels are shown weekly.

Figure 13: Same as Figure 2 but for 700-dbar float 680, Eddy C. Note that the oscillations in the time series of background temperature result from the north-south motion of the float in the presence of a meridional mean background temperature gradient.

Figure 14: Same as Figure 4 but for float 680, Eddy C.

Figure 15: Time series of TOA from sound source D as recorded by float 680, showing continuous oscillatory motion starting with the formation of Eddy C around 27 August 2006.

Figure 16:  Same as Figure 7 but for the time period of Eddy C. Trajectory segments from three different 700-m floats are shown: 680 (dot), 582 (x) and 586 (star). The first seven panels show the eddy formation at the TGB, while the last five panels illustrate the cessation of float looping when Eddy C hit the Gulf Stream. There is a 2.5-month gap between the two sequences.

Figure 17:  Time series of temperature for the three floats in Eddy C: 680 (blue), 582 (red) and 586 (black).

Figure 18:  (a) Contoured potential temperature based on data at nominal depths of 1000, 1200, and 1500 m (mean depths 1021, 1233, and 1517 m) from the BIO-50W East site for 23 November 1978 - 19 January 1979. Selected contours of $\sigma_{15}$ (grey curves) are also shown. Note that the $\sigma_{15}$ contours are based on data from only the nominal 1000 and 1500 m depths because of high noise levels in the 1200 m salinity data. Nearly identical minima in potential temperature of 3.6°C occurred at 1500 m at 1800 01 January 1979 and 1800 11 January 1979 (vertical lines). Minima in potential temperature of 4.6°C and 3.8°C occurred at 1200 m at 1800 01 January 1979 and 0000 12 January 1979 respectively. Estimated mean measurement depths are marked with horizontal lines. (b) Similar contoured potential temperature at the East site for 22 July 1979 – 05 August 1979. A minimum in potential temperature of 2.3°C occurred at the 1000 m level at 0000 29 July 1979 (vertical line). A minimum in potential temperature of 4.4°C occurred at 1200 m at the same time. (c) Contoured potential temperature based on data at nominal 1000 and 1500 m depths (mean depths 1017 and 1513 m) from the South site for the same period as the middle panel. A minimum in potential temperature of 6.9°C occurred at 1000 m at 0600 01 Aug 1979

(vertical line). The results are blanked in the 1100–1400 m depth range to emphasize that the vertical interpolation is based on data from only 1000 m and 1500 m depths.

Figure 19. (a) Potential temperature − salinity diagram showing 1500 m observations from the BIO-50W East 1a,b analysis period and 1000m observations for the East 2 and South 2 analysis periods. Contours of $\sigma_{15}$ are shown at 0.2 kg/m$^3$ intervals with labels at 1 kg/m3 intervals. A reference curve (solid line) is derived from the annual HydroBase climatology interpolated to the location of the East site. Hudson 89037 50°W CTD data (blue dots) and end points (large open circles) of mixing lines (dashed curves) discussed in the text are also shown. (b) Scatter plot of $\sigma_{15}$ vs. salinity anomaly relative to the reference curve for the same points plotted in the Figure 19a. The first data point in each of the two analysis periods is highlighted with a black circle. The salinity associated with the 1.8°C minimum reference potential temperature is about 34.89. Since this value is also used as the reference salinity for all potential temperatures less than 1.8°C, the curvature of the mixing lines is different for potential temperatures below and above 1.8°C. This breakpoint occurs at salinity anomalies of -0.86 for the East 2 mixing curve and -1.03 for the South 2 mixing curve.

Figure 20: Data and model fits for the 1000 m BIO-50W East 2 analysis period. (a) Estimated eddy trajectory oriented towards 205°T. Dots along the trajectory corresponding to the 31 data points at 6 hour intervals used to fit the parametric model. The estimated translation speed was 0.10 m s$^{-1}$ and 6 hours corresponds to a 2.2 km distance increment. Also shown are selected flow anomaly vectors from the fitted eddy model at the point of closest approach (open triangle) at distances of multiples of R/2 from the eddy center. (b) Range from the mooring to the eddy

center as a function of time. (c) Bearing from the mooring to the eddy center. (d) Measured (thick grey curve with dots) and modeled (thin black curve) composite u-component. Residuals (observed - fit) are shown as a dotted curve. (e) Measured and modeled composite v-component and residuals. (f) Measured and modeled composite longitudinal flow ut (towards 205°T) and residuals. (g) Measured and modeled transverse flow vt (towards 115°T) and residuals. (h) Measured and modeled composite rate. (i) Measured and modeled temperature anomaly and residuals. The model temperature anomaly is simply the least-squares projection of the normalized Gaussian streamfunction onto the measured temperature anomalies.

Figure 21: Data and model fits for the 1000 m BIO-50W South 2 analysis period as in Figure 20. The analysis period includes 14 points at 6 hour intervals. (a) Measured (thick grey curve with dots) and modeled (thin black curve) composite longitudinal flow ut (towards 224°T). The flow reversal to a minimum composite longitudinal flow of −0.20 m s$^{-1}$ corresponds to a peak longitudinal flow anomaly of about 0.35 m s$^{-1}$ since the estimated translation speed was 0.16 m s$^{-1}$. (b) Measured and modeled composite transverse flow vt (towards 134°T) and residuals. (c) Measured and modeled temperature anomaly and residuals.

Figure 22: Stick plots of horizontal flow at the 1000 m levels for 22 July 1979 – 05 August 1979 for the BIO-50W East site (upper panel) and the South site (lower panel). A common scale arrow is shown in the upper panel. The flows are partitioned into model anomalies (thin black lines) and residual (thicker grey lines) defined as the observed flow minus the model anomaly. The two plots use different upwards directions as indicated by the North arrows. The upwards flow direction for the upper panel (East site) is towards the 044°T transverse flow direction. The upwards flow direction for the lower panel (south site) is towards the 294°T longitudinal flow direction. Vertical lines mark the times of closest approach from the fits at the two sites. The two sites are separated by 50 km and the difference in the times of closest approach is 80 hours, giving a mean advection speed of 15 km/d or 0.17 m s$^{-1}$ towards 294°T if the two anomalies are related to a single feature. For comparison, the background flows from the model fits at the East and South sites were 0.10 m s$^{-1}$ towards 205°T and 0.16 m s$^{-1}$ towards 224°T respectively.

Figure 23: Schematic illustration of an SCV subducted downward along sloping isopycnals that represent the Gulf Stream. Dashed lines show how an isobaric 700-m float could "pop" out the top of the eddy and a 1500-m float would move to a vertical position higher in the eddy after the eddy has crossed the Gulf Stream.

Figure 24: Expanded view of the bathymetry of the eastern slope of the Grand Banks. Contour interval is 500 m, highlighted every 1000 m. The positions one rotation period after the three eddies were first detected by the wavelet analysis of the RAFOS float trajectories are indicated by colored circles with radii of float looping.

Figure 25: Same as Figure 24 but showing a close-up of the bathymetry of the TGB and the trajectories of each of the three floats that were used to identify Eddies A, B and C. Small open circles indicate float position every day, and large circles represent the actual looping radius of the floats in each eddy (664 in Eddy A, 581 in Eddy B and 680 in Eddy C). Isobaths are drawn every 200 m and in bold every 1000 m.

Figure A1:  A synthetic eddy signal, and its recovery. Panel (a) shows the eastward (solid black) and northward (dashed black) components of a circular, anticyclonically rotating displacement signal. The geometric mean velocity V(t) in (b) and geometric mean radius R(t) in (c) linearly increase in magnitude, while the period $2\pi/|\omega|$ in (d) remains constant. In each of panels (b—d), the recovered signals using the method described in the text are shown with dashed gray lines for time periods before t=5 days and after t=45 days, and as solid gray lines during other times. The solid gray lines are generally not visible since they are exactly overlapped by the true signals.

Figure A2:  The radius/velocity plot for the synthetic eddy signal shown in Figure A1. As in Figure A1, the solid black curve shows the true properties, while the solid gray and dashed gray lines are the recovered properties from the central time period and the "edge effect" time periods, respectively.

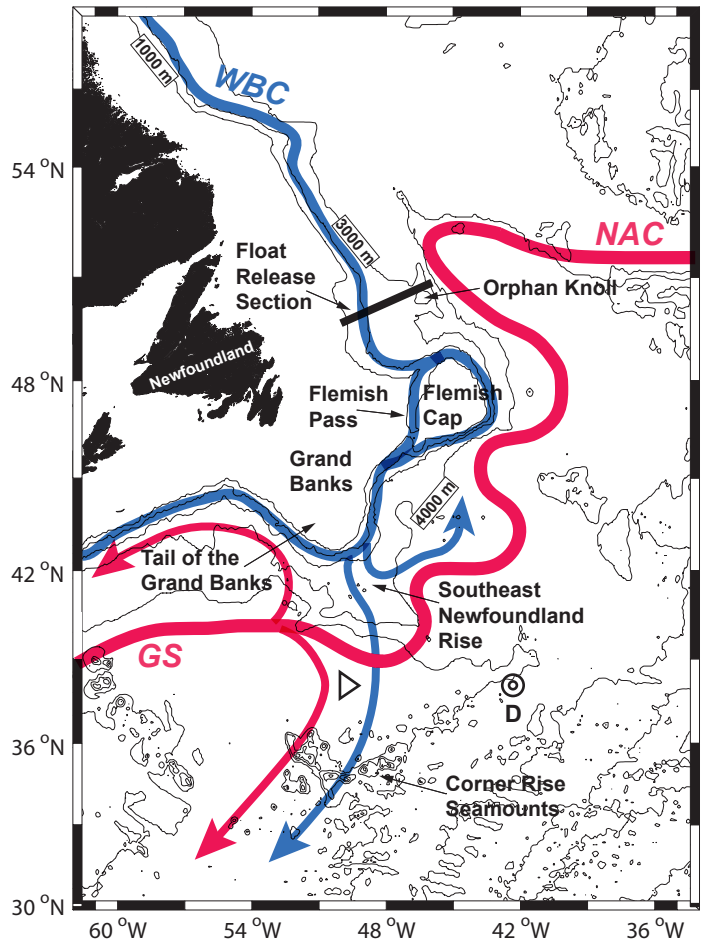

Figure 1

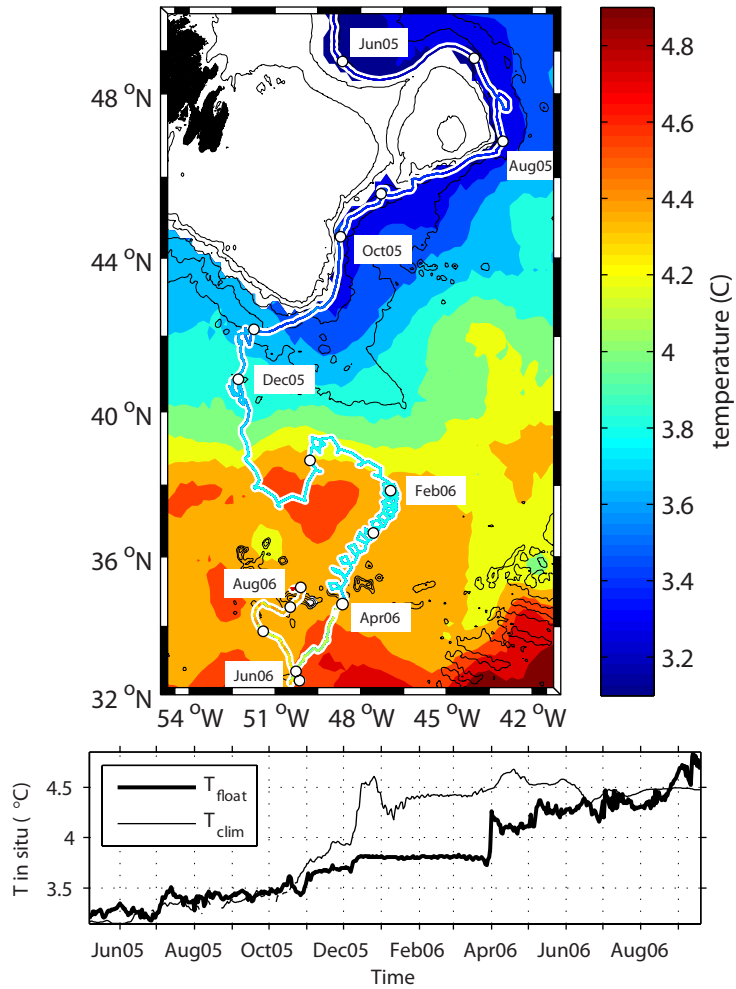

Figure 2

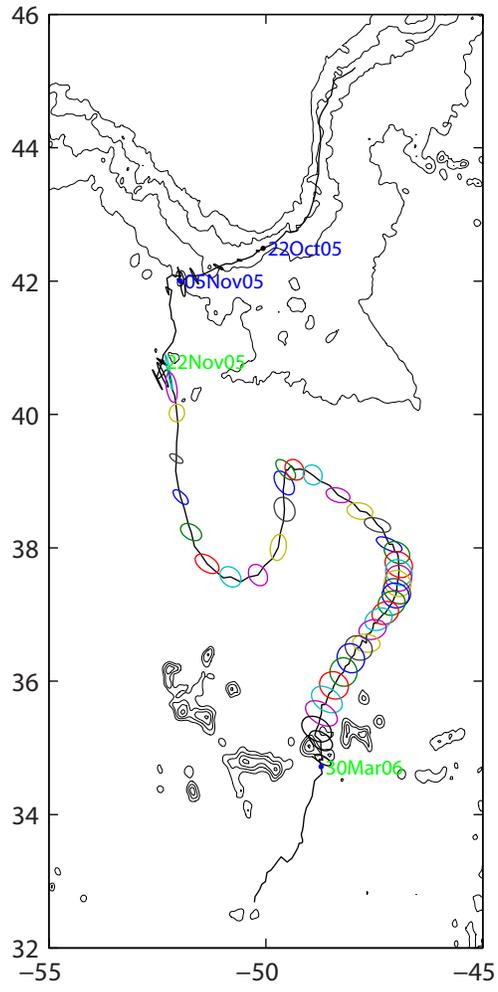

Figure 3

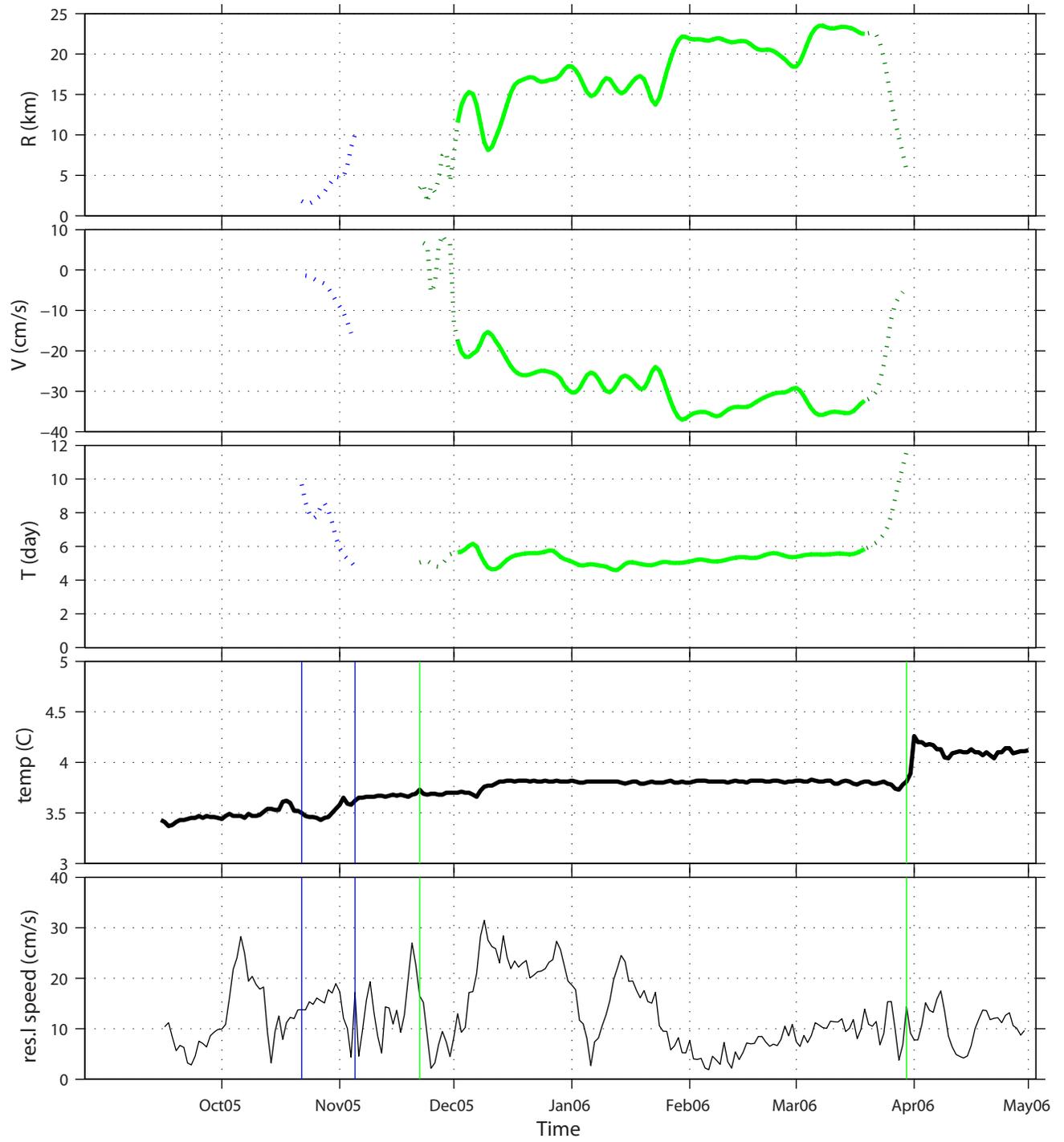

Figure 4

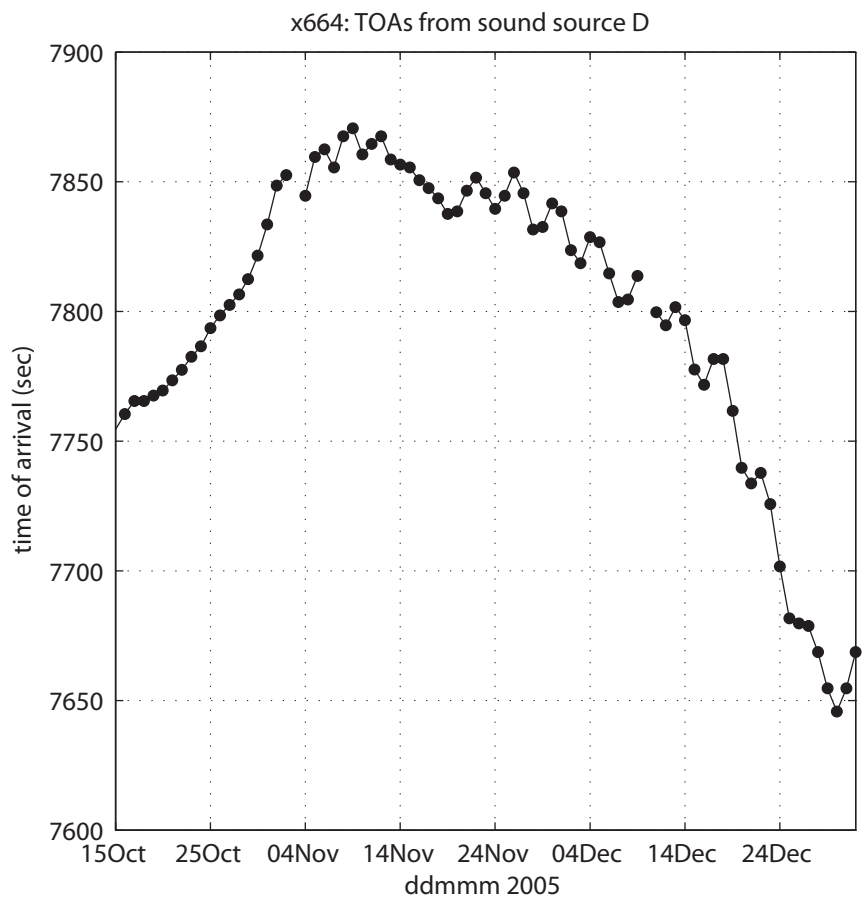

Figure 5

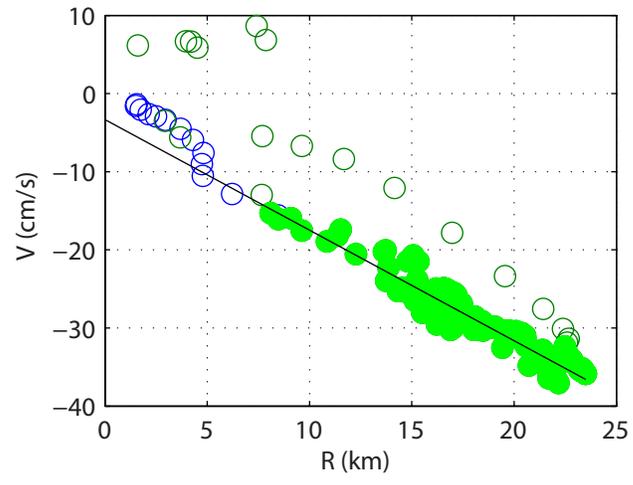

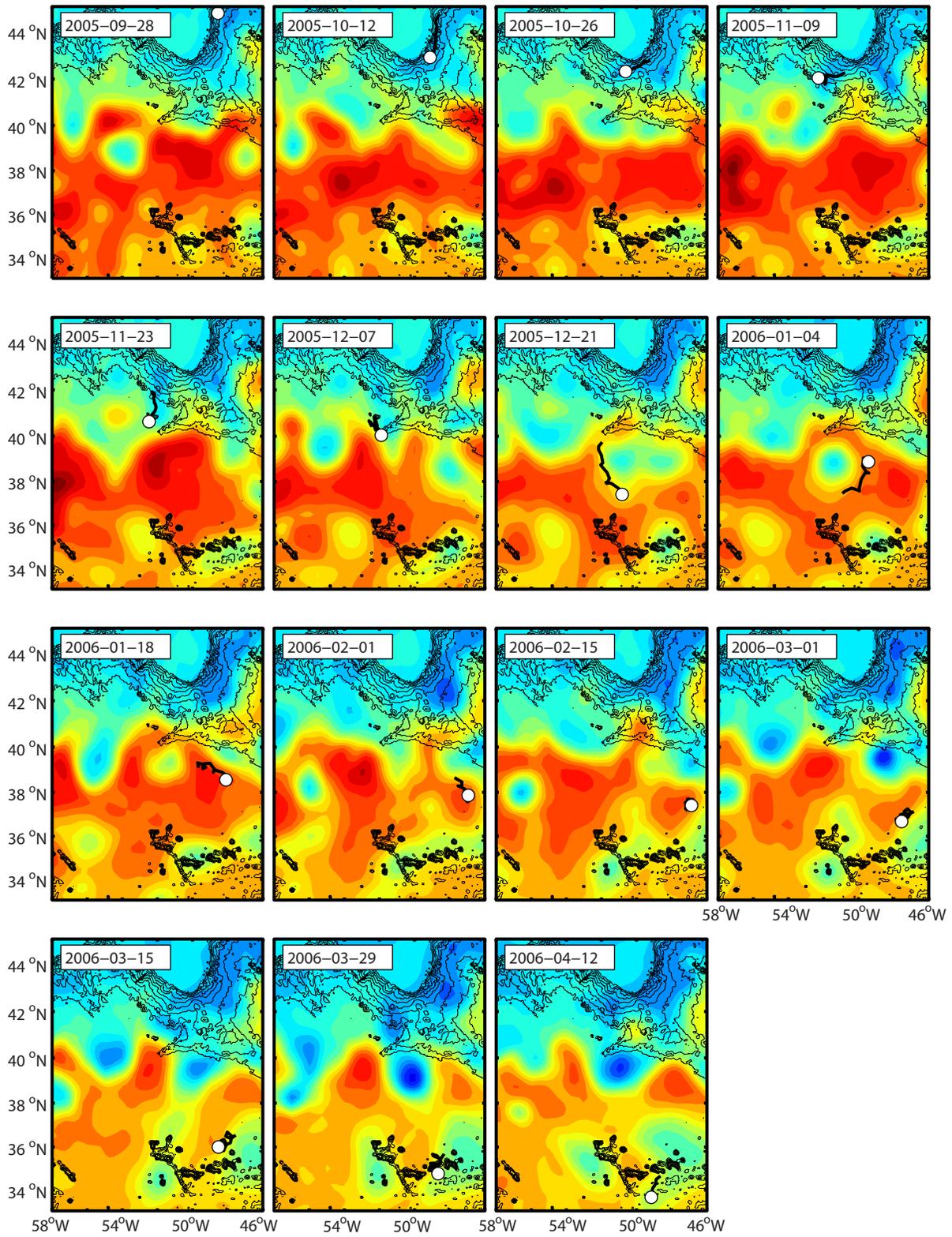

Figure 7

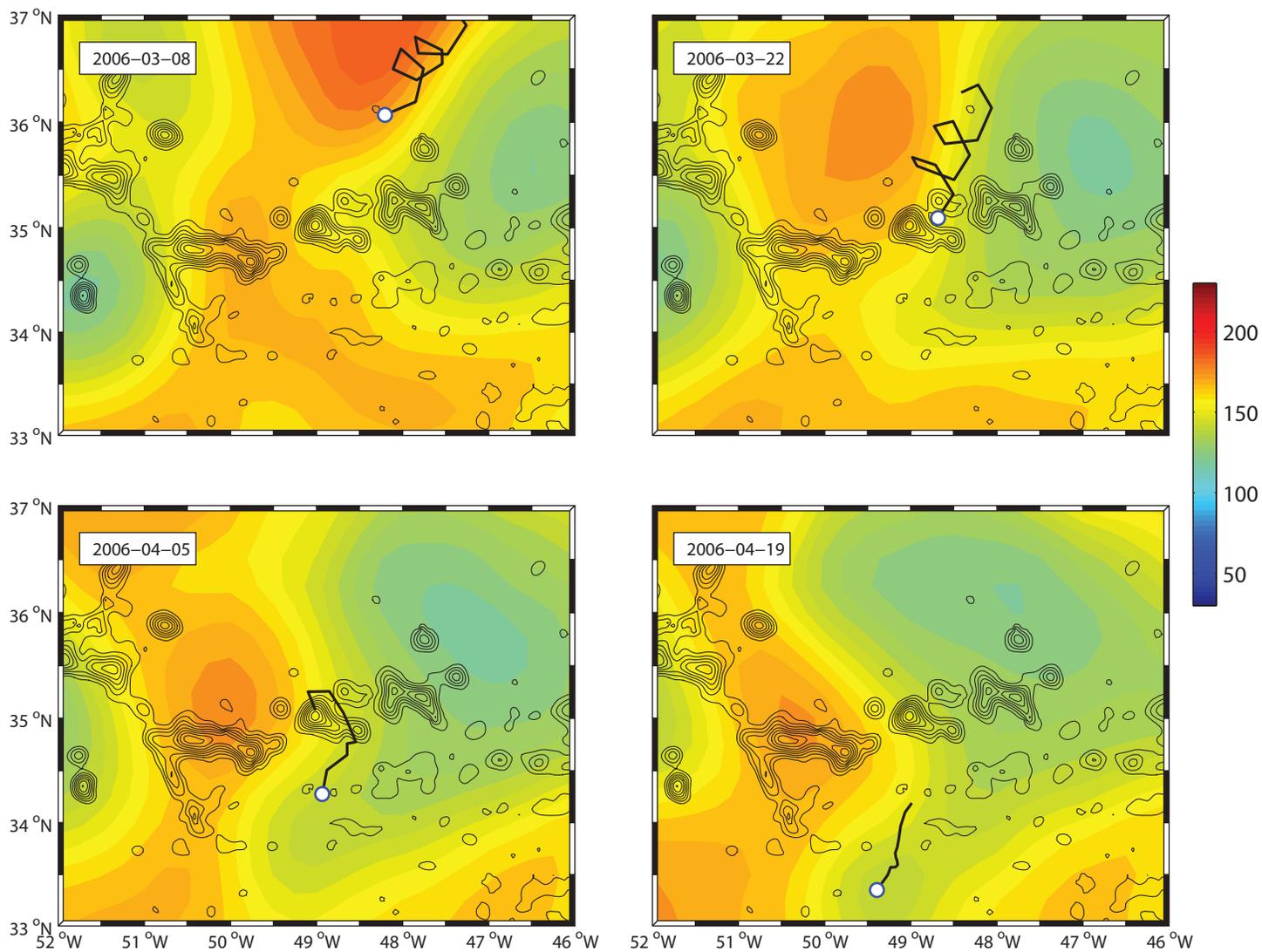

Figure 8

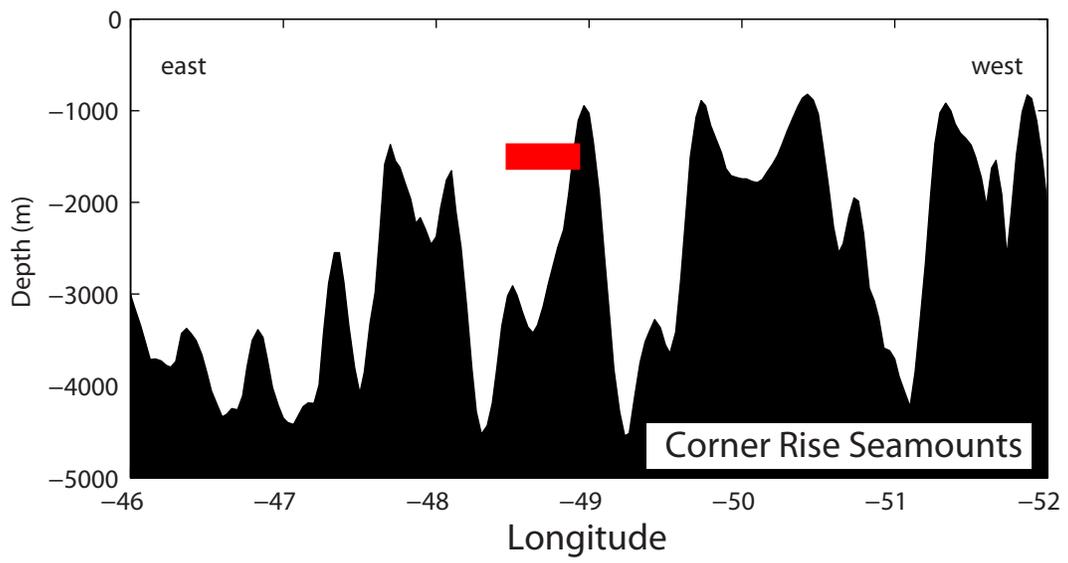

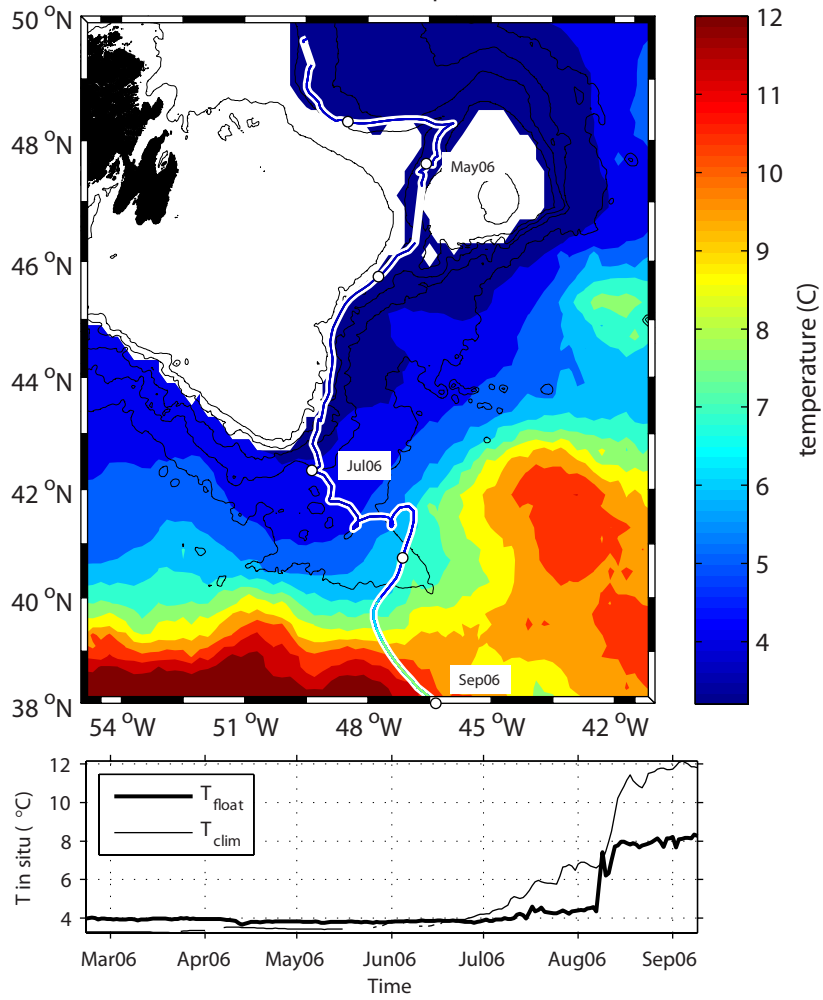

Float path and temperature for 581
and in situ annual mean temperature at 700m

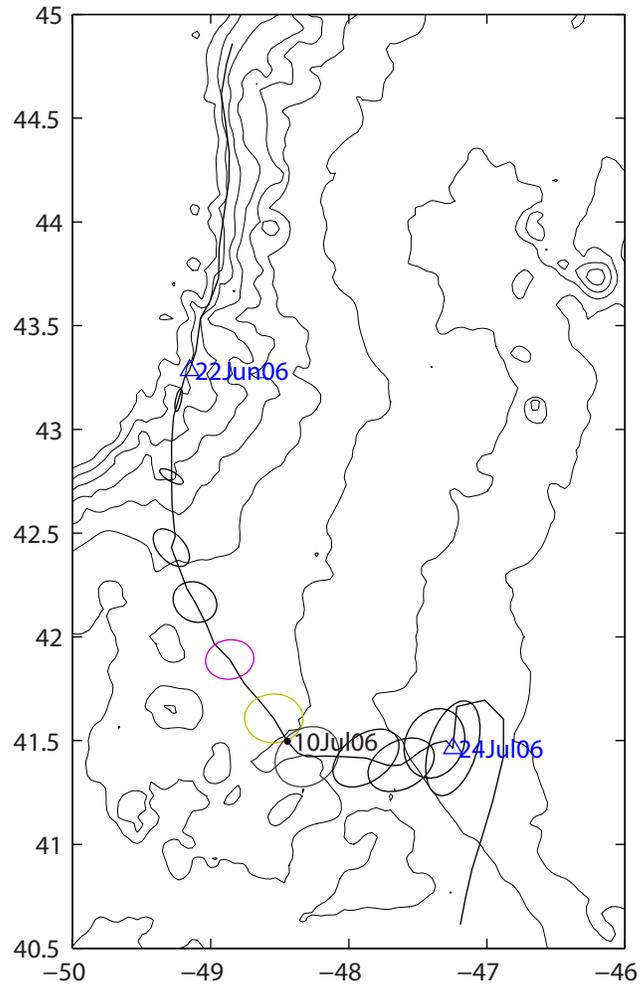

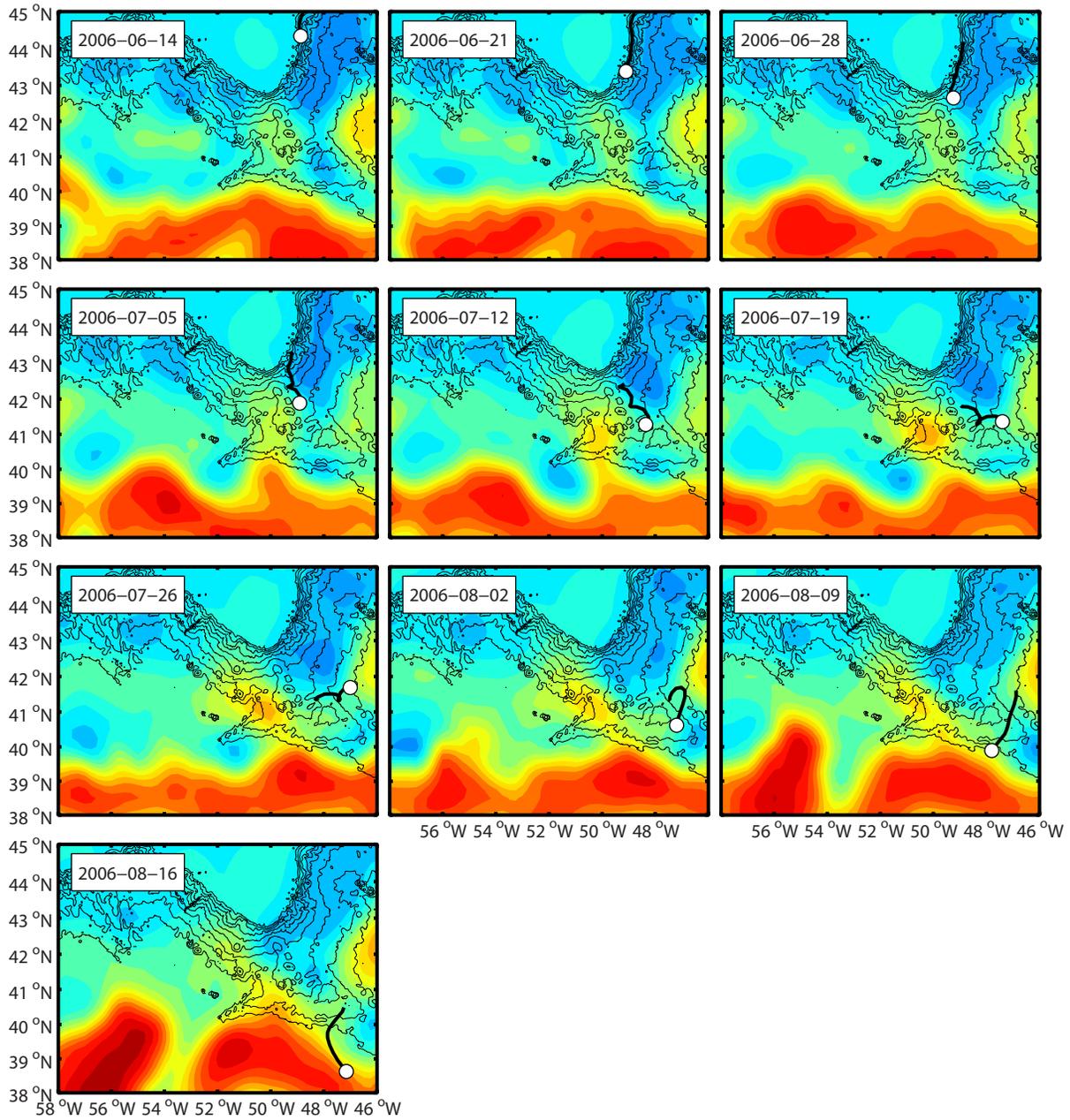

Figure 12

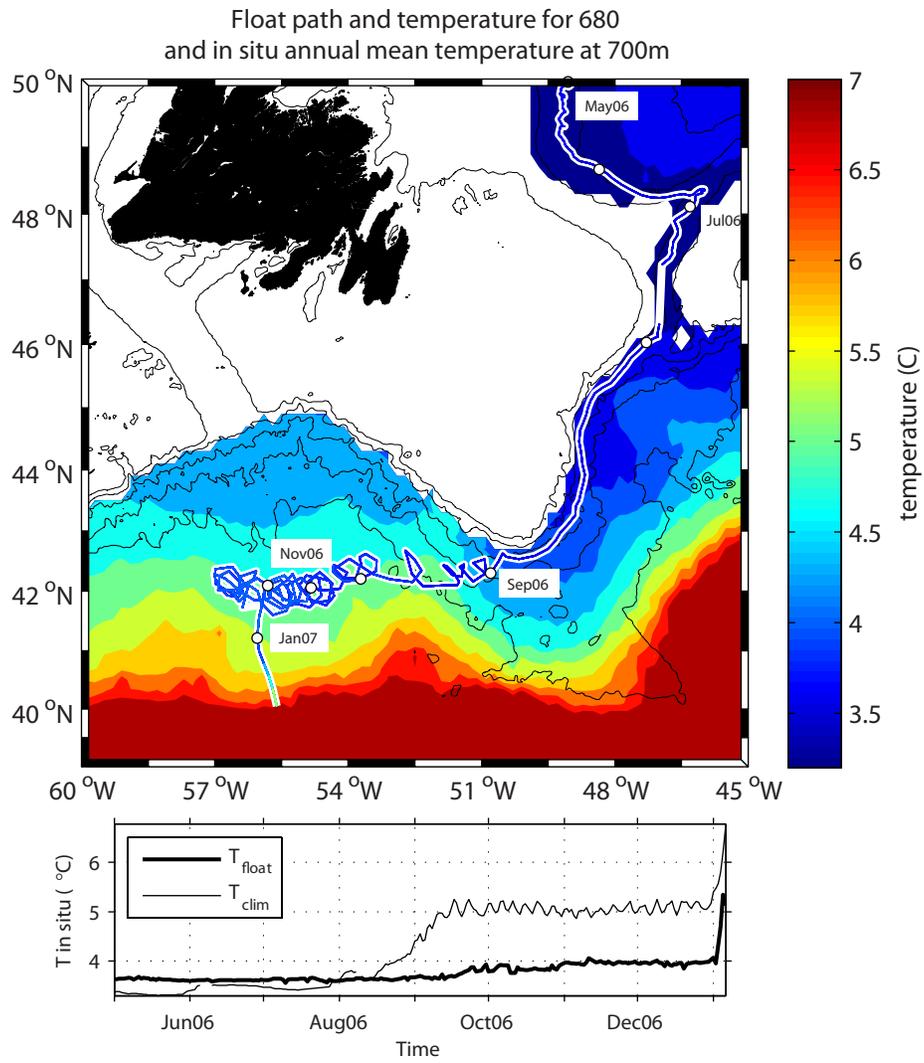

Float path and temperature for 680
and in situ annual mean temperature at 700m

Figure 13

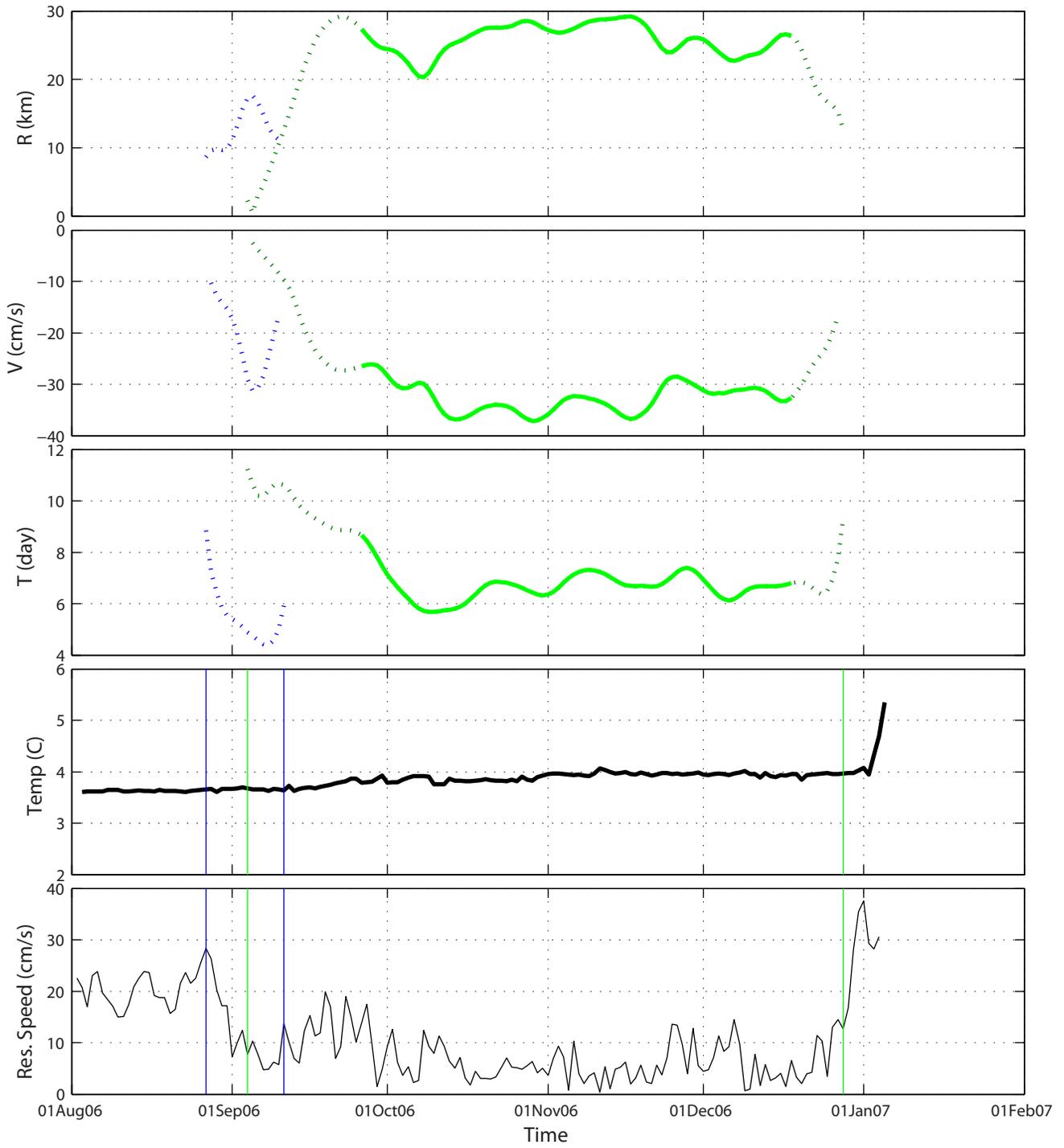

Figure 14

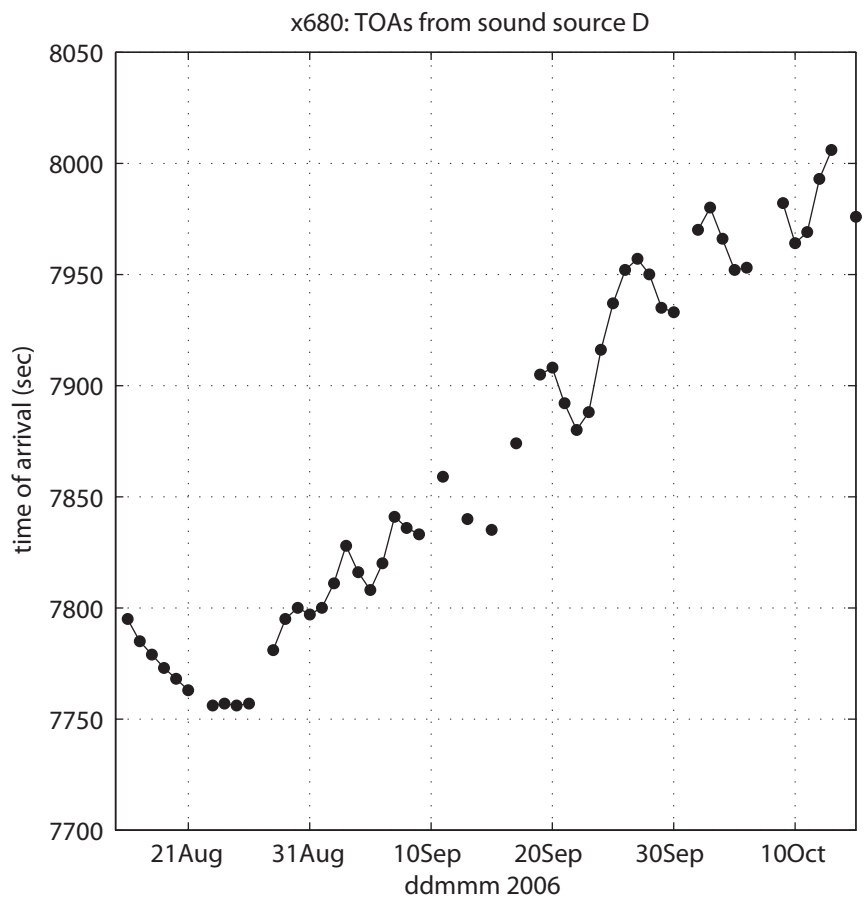

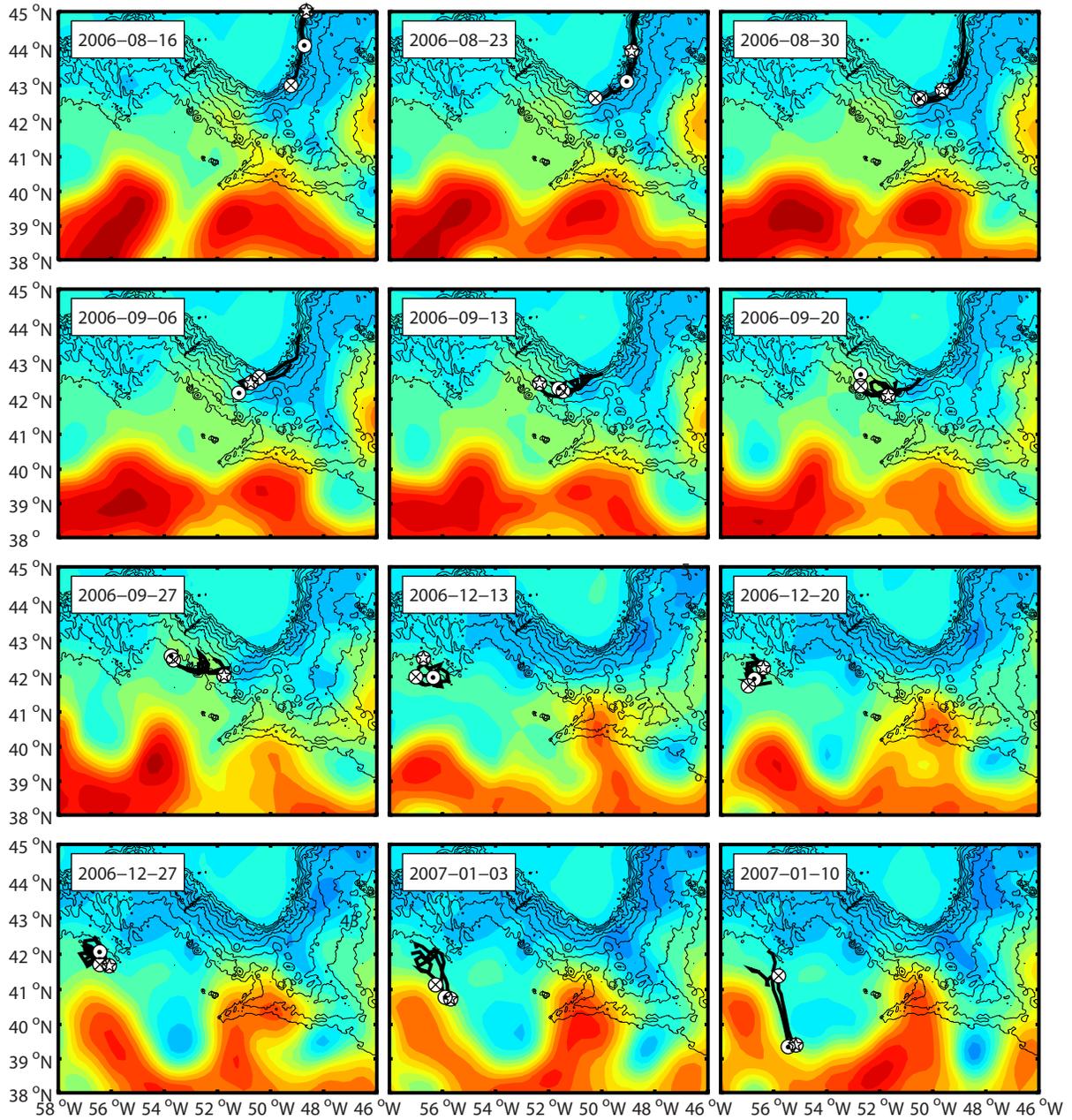

Figure 16

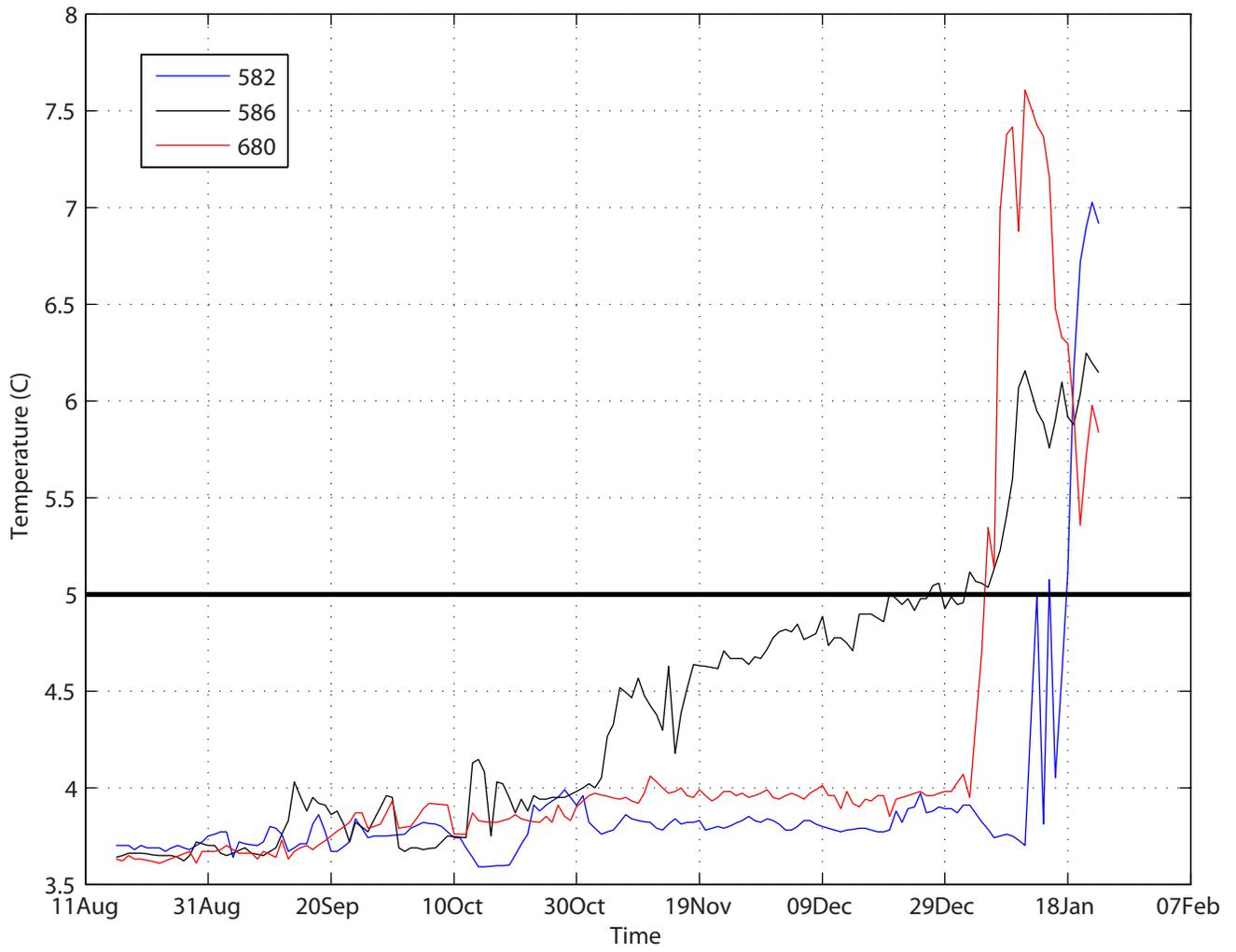

Figure 17

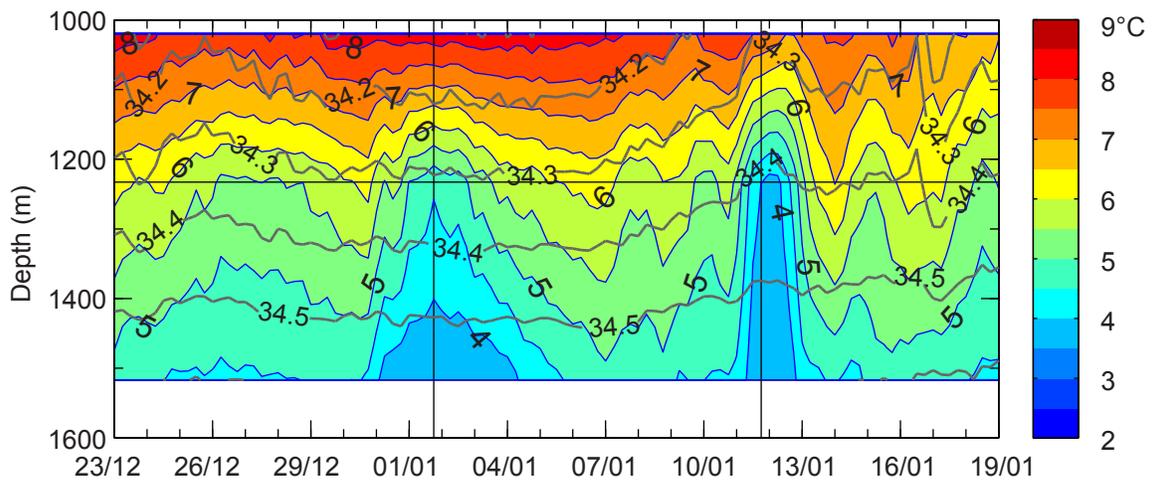

Figure 18a

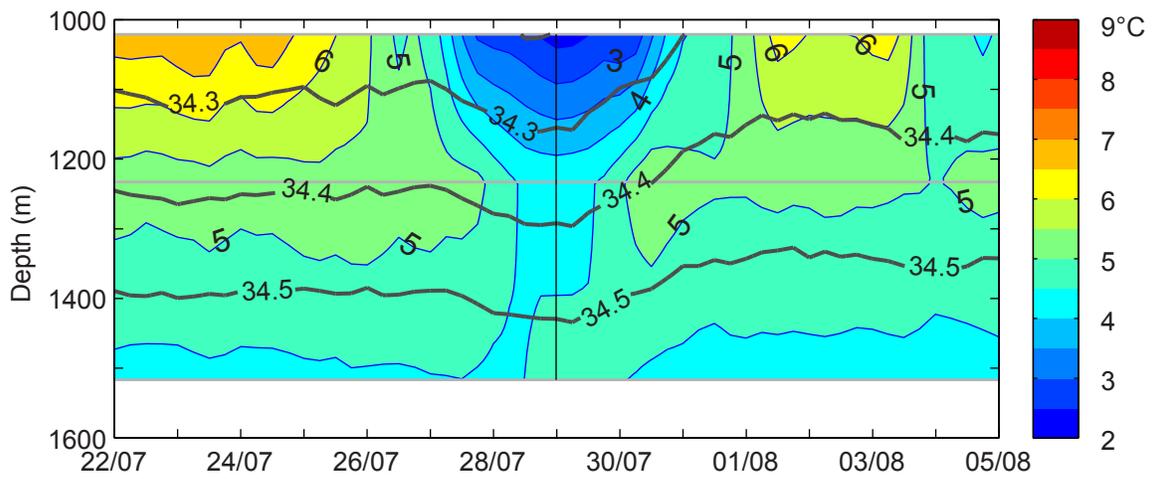

Figure 18b

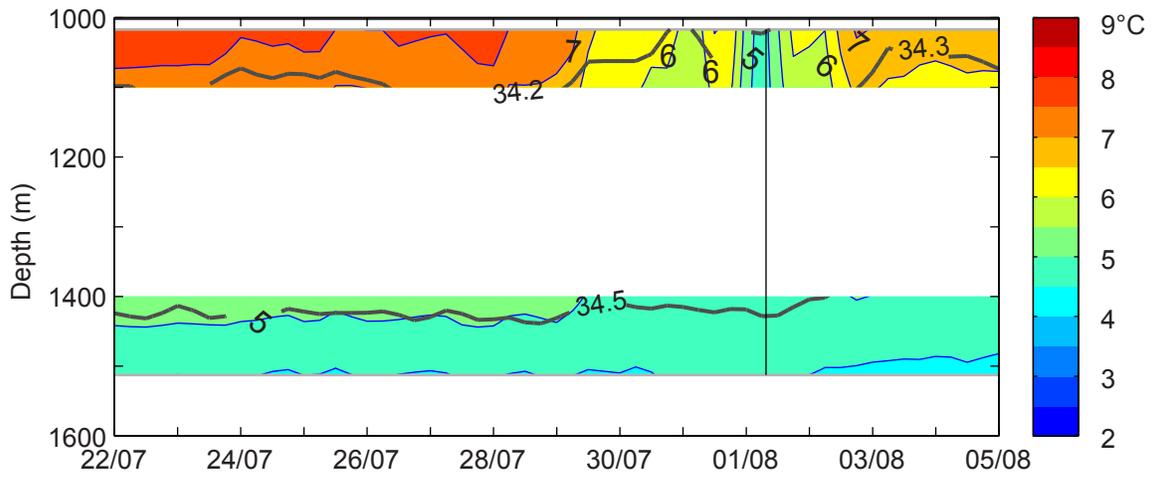

Figure 18c

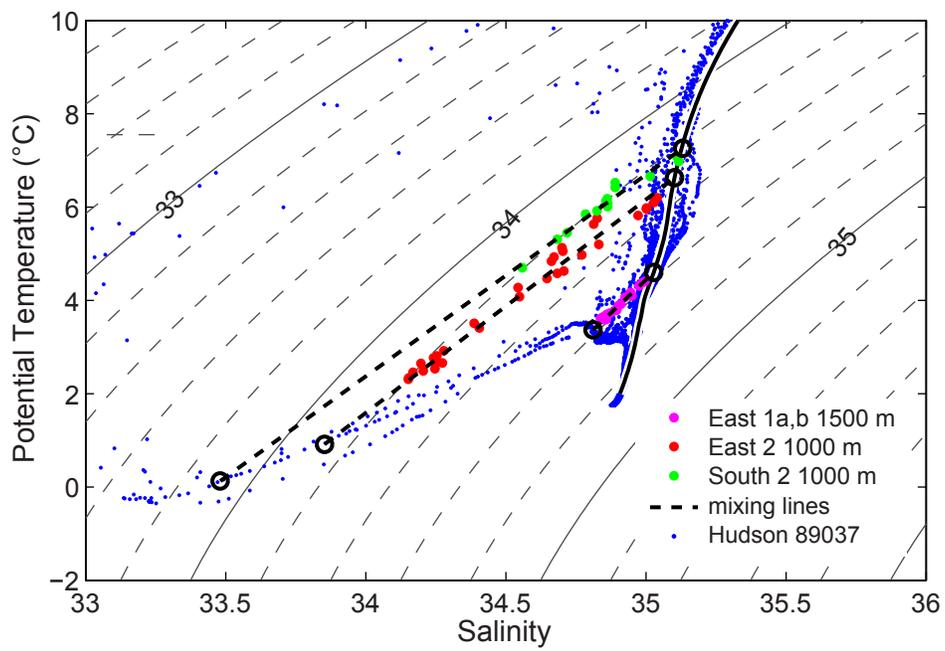

Figure 19a

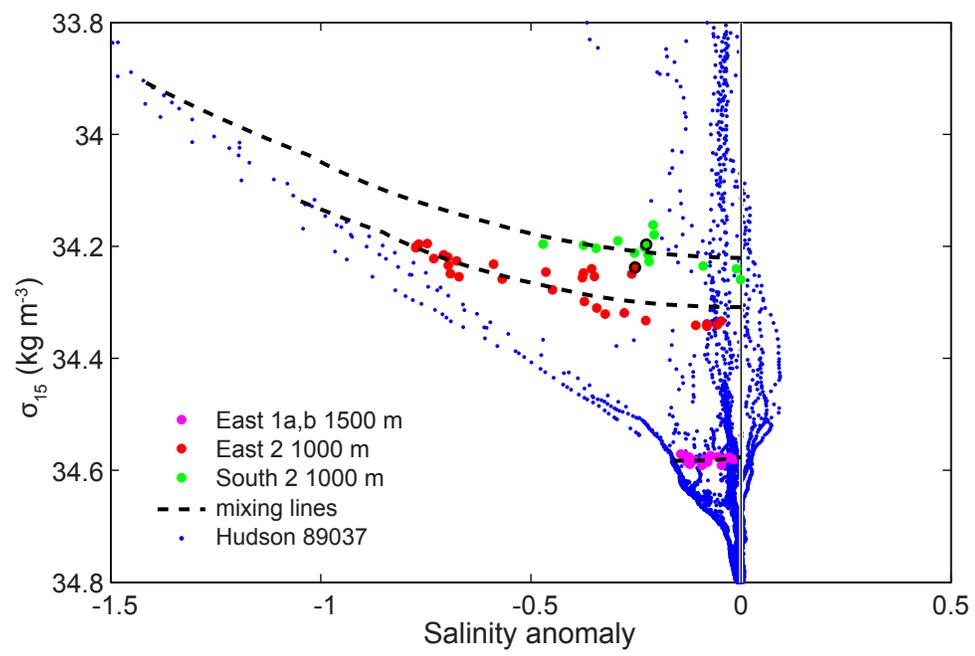

Figure 19b

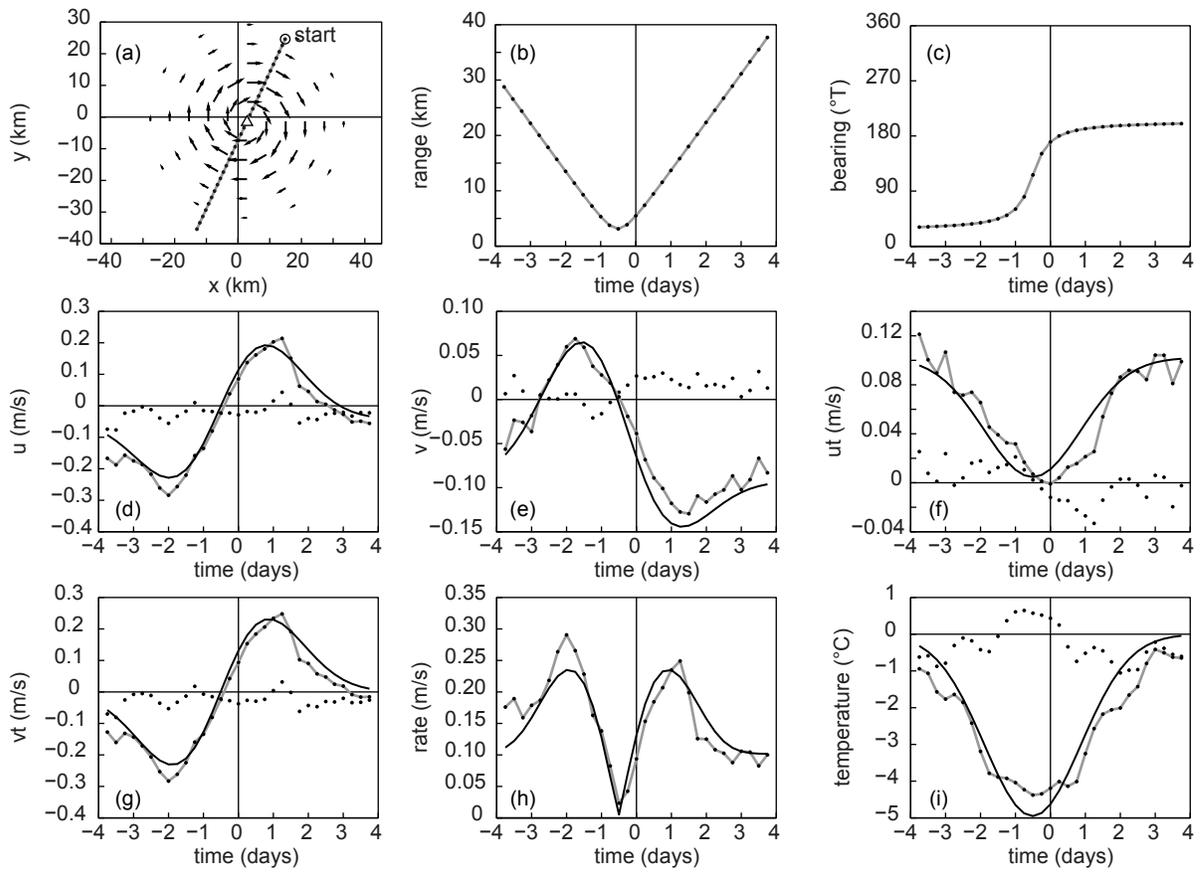

Figure 20

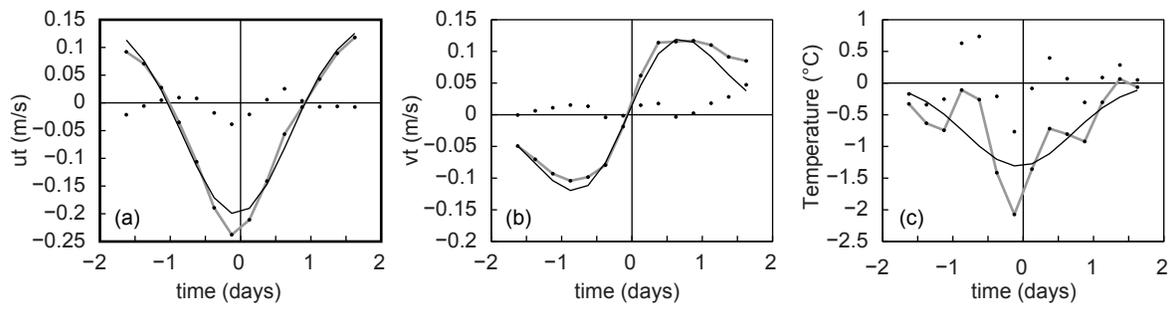

Figure 21

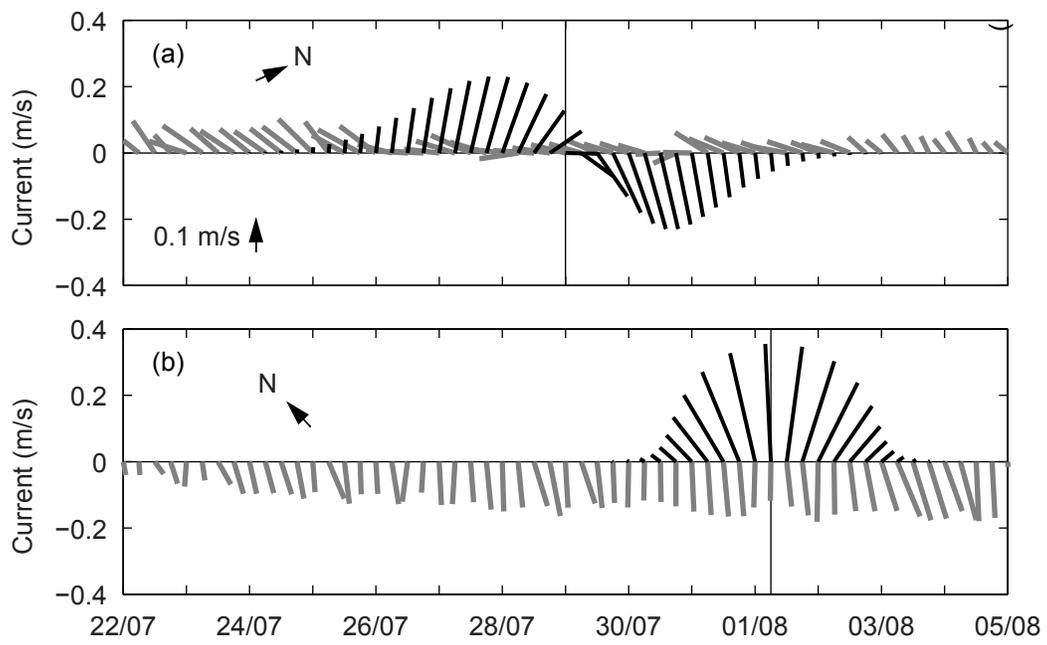

Figure 22

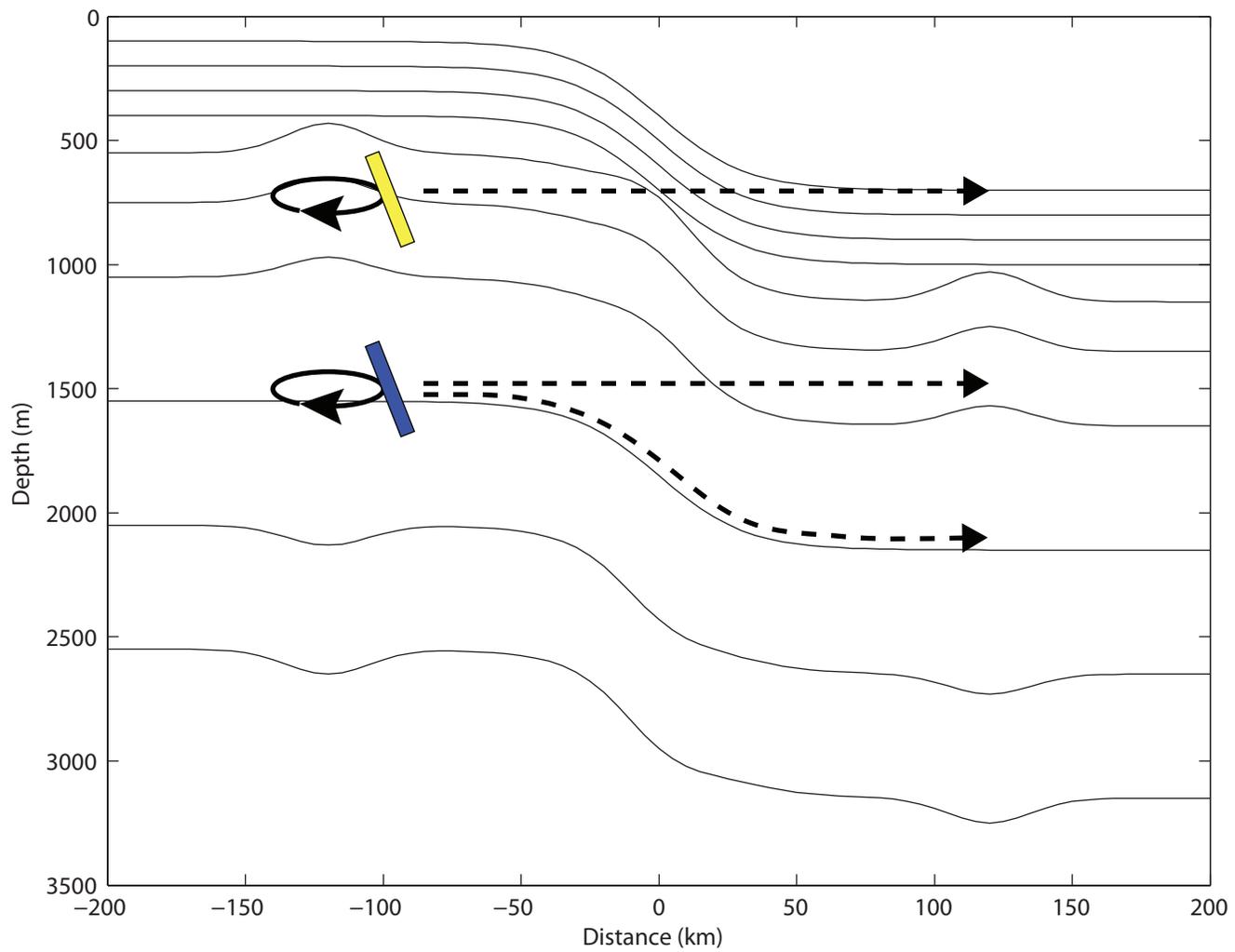

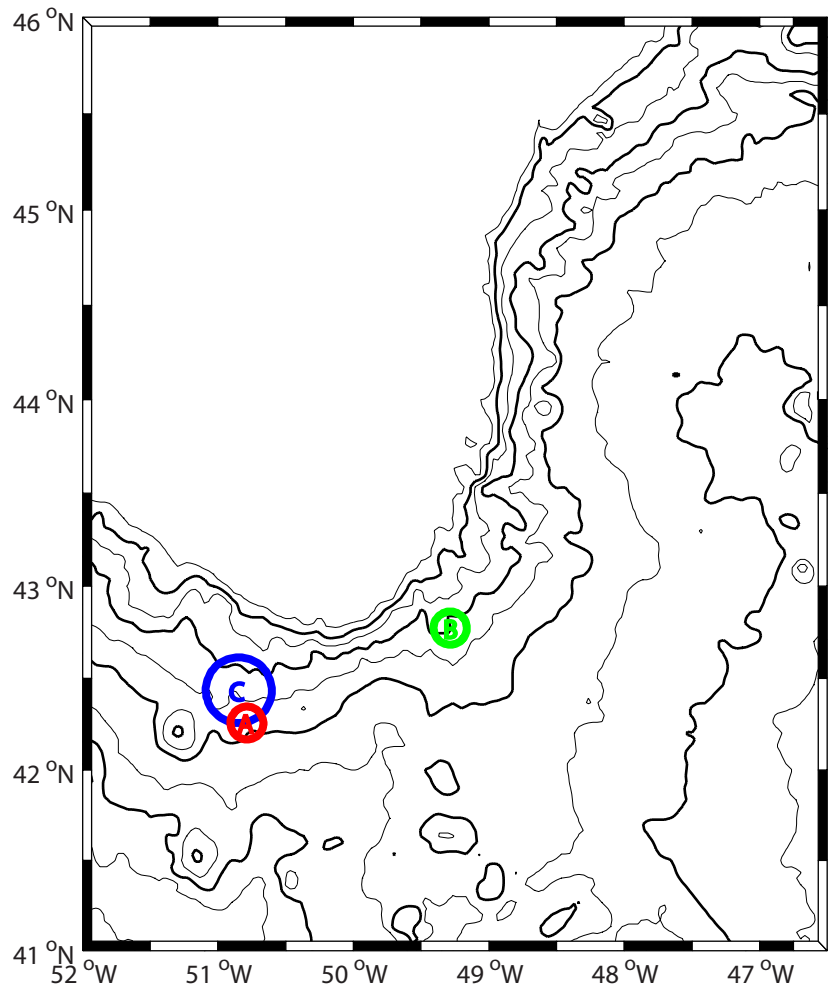

Figure 24

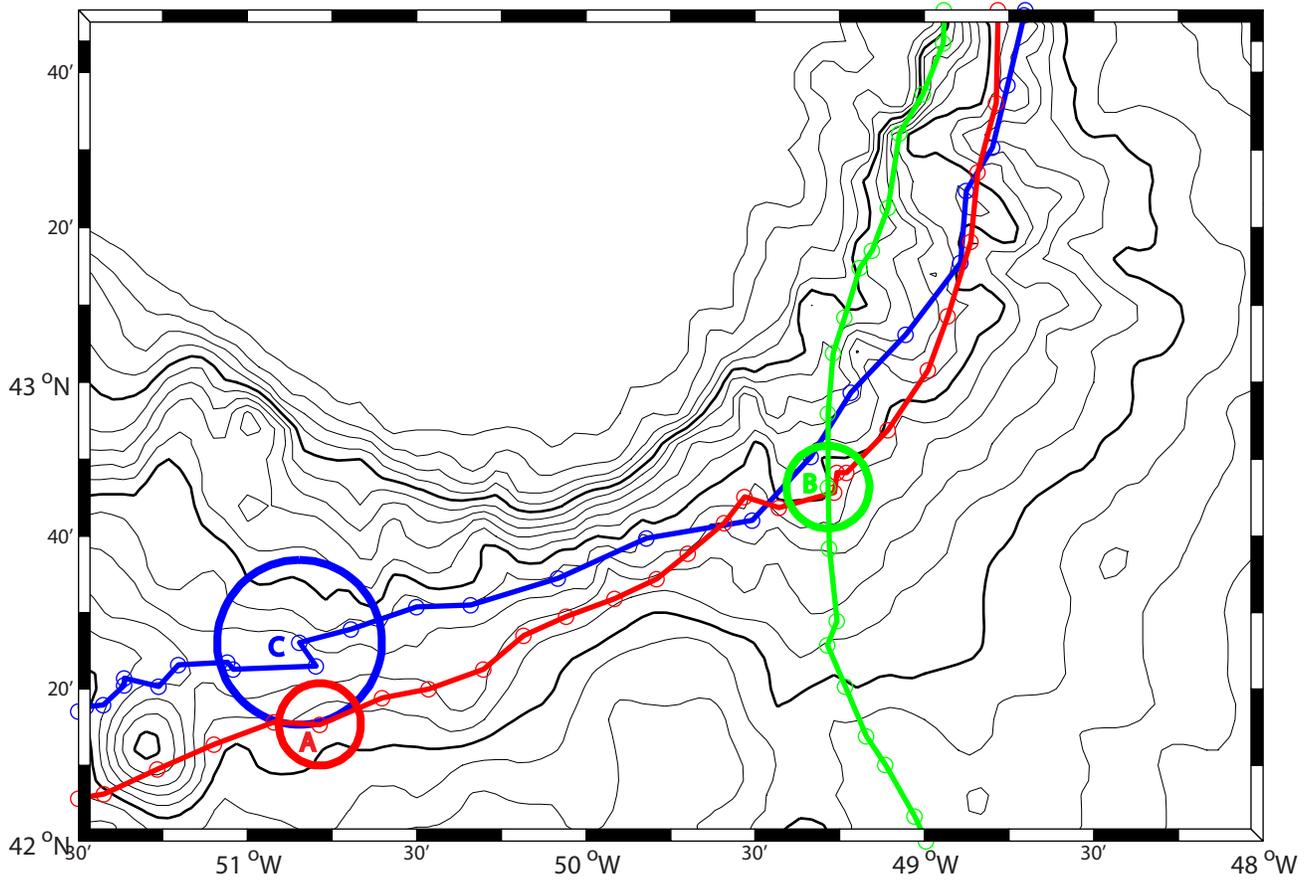

Figure 25

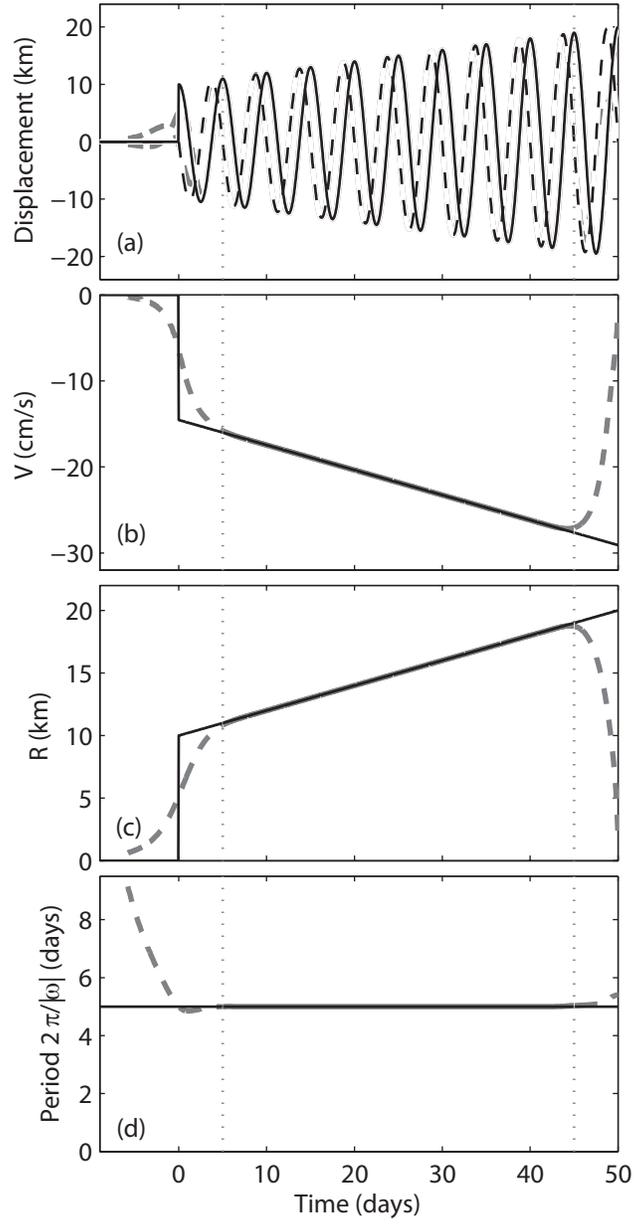

Figure A1

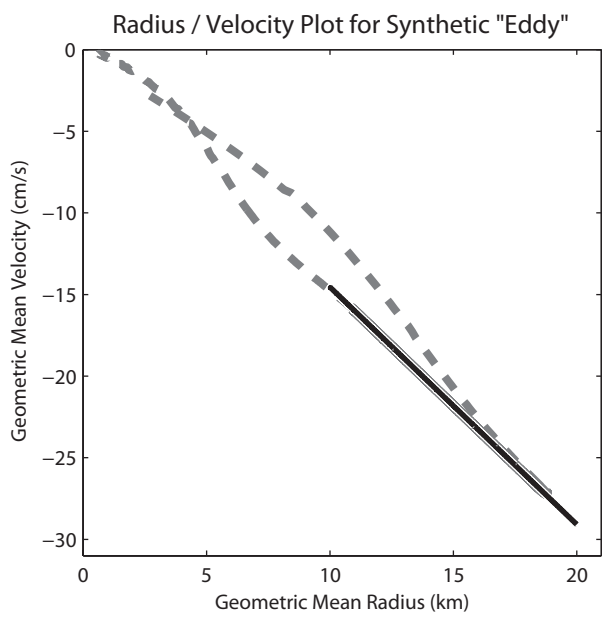